\documentclass[twoside,twocolumn,english,aps,showpacs,prl,superscriptaddress]{revtex4-1}
\usepackage[T1]{fontenc}
\usepackage[latin9]{inputenc}
\usepackage{color}
\usepackage{bm}
\usepackage{enumerate}
\usepackage{subfigure}
\usepackage{amsmath}
\usepackage{amsthm}
\usepackage{amssymb}
\usepackage{graphicx}
\usepackage{esint}
\usepackage{mathrsfs}
\usepackage{booktabs}
\usepackage{multirow}
\usepackage{makecell}
\usepackage{braket}
\usepackage{babel}
\usepackage{extarrows}
\usepackage{color}
\usepackage[plainpages = false, pdfpagelabels, 
                 bookmarks,
                 bookmarksopen = true,
                 bookmarksnumbered = true,
                 breaklinks = true,
                 linktocpage,
                 colorlinks = true,
                 linkcolor = blue,
                 urlcolor  = blue,
                 citecolor = blue,
                 anchorcolor = green,
                 hyperindex = true,
                 hyperfigures
                 ]{hyperref} 
\usepackage{dcolumn}   

\makeatletter

\makeatletter
\newcommand*{\rom}[1]{\expandafter\@slowromancap\romannumeral #1@}
\makeatother

\usepackage{times}{\Large }

	\renewcommand{\Re}{\operatorname{Re}}

	\DeclareMathOperator{\tr}{tr}  		

	\DeclareMathAlphabet{\mathbbold}{U}{bbold}{m}{n}

    \renewcommand{\d}{\mathrm{d}}

\newtheorem{theorem}{Theorem}

\newtheorem{lemma}{Lemma}

\theoremstyle{definition}

\theoremstyle{remark}
\newtheorem*{remark}{Remark}

\makeatother

\begin{document}

\title{Extracting Error Thresholds through the Framework of Approximate Quantum Error Correction Condition}

\author{Yuanchen Zhao}
\affiliation{State Key Laboratory of Low Dimensional Quantum Physics, Department of Physics, Tsinghua University, Beijing, 100084, China}
\affiliation{Frontier Science Center for Quantum Information, Beijing 100184, China}

\author{Dong E. Liu}
\email{Corresponding to: dongeliu@mail.tsinghua.edu.cn}
\affiliation{State Key Laboratory of Low Dimensional Quantum Physics, Department of Physics, Tsinghua University, Beijing, 100084, China}
\affiliation{Beijing Academy of Quantum Information Sciences, Beijing 100193, China}
\affiliation{Frontier Science Center for Quantum Information, Beijing 100184, China}
\affiliation{Hefei National Laboratory, Hefei 230088, China}

\begin{abstract}
The robustness of quantum memory against physical noises is measured by two methods: the exact and approximate quantum error correction (QEC) conditions for error recoverability, and the decoder-dependent error threshold which assesses if the logical error rate diminishes with system size. Here we unravel their relations and propose a unified framework to extract an intrinsic error threshold from the approximate QEC condition, which could upper bound other decoder-dependent error thresholds. Our proof establishes that relative entropy, effectively measuring deviations from exact QEC conditions, serves as the order parameter delineating the transition from asymptotic recoverability to unrecoverability. Consequently, we establish a unified framework for determining the error threshold across both exact and approximate QEC codes, addressing errors originating from noise channels as well as those from code space imperfections. This result sharpens our comprehension of error thresholds across diverse QEC codes and error models. 
\end{abstract}


\date{\today}

\maketitle

\section{Introduction}
Quantum Error Correction (QEC) is crucial for fault-tolerant quantum computation \cite{shor,stean,Calderbank96,gottesman1997stabilizer,ref:dennis,KITAEV20032}. At its core, the QEC or Knill-Laflamme condition \cite{knillTheoryQuantumErrorcorrecting1997} is important for accurately recovering quantum information from error channels. However, a fundamental conflict exists between the exact QEC condition, aimed at exact recoverability, and the error threshold theorem \cite{knill2,aliferis,ref:dennis,colorcode,fowler,Bombin-arxiv,Kovalev_sublinear,gottesmanFaultTolerantQuantumComputation2014,breuckmannConstructionsNoiseThreshold2016,kubica_three-dimensional_2018,kovalevNumericalAnalyticalBounds2018,Vuillot,Pryadko2020maximumlikelihood,ref:flammia,song_optimal_2022}, which asserts that keeping error rates below a critical threshold significantly mitigates logical errors. Real-world systems \cite{NiggScience14,Ofek16Nature,Hu19NP,Andersen20NP,Erhard21Nature,GoogleAI21Nature,Luo21PNAS,Ryan-Anderson21PRX,Marques22NP,ZhaoPRL-22,Egan20arXiv} (described by many realistic noise models \cite{aliferis,ref:dennis,colorcode,fowler,Bombin-arxiv,breuckmannConstructionsNoiseThreshold2016,kubica_three-dimensional_2018,kovalevNumericalAnalyticalBounds2018,Vuillot,flammia_compare,darmawan_tensor,yangQPRA2022} ) often fail to satisfy QEC condition, highlighting an intrinsic nonzero logical error rate in finite systems. This discrepancy has led to the adoption of error thresholds.

Within the notion of mixed state phase transition (see Refs.~\cite{leeQuantumCriticalityDecoherence2023,bao2023mixedstate,sang2023mixedstate,dai2023steadystate}), Ref.~\cite{ashvin} investigated a specific example of toric code with single-qubit Pauli noise by linking the threshold to singular behaviors of several intrinsic characteristics without the knowledge of particular decoder algorithm.
However, it still lacks a mathematically rigorous proof and raises questions about the validity of the intrinsic characterization in general cases. 
This prompted us to reconsider the discrepancy between the QEC condition and error threshold, further inspiring a novel idea: Can we reconcile the code-specific QEC condition with the decoder-dependent thresholds in advancing practical QEC strategies against diverse noise types in quantum memory?

We consider the Approximate QEC (AQEC) condition \cite{schumacher_approximate_2001,tyson_two-sided_2010,beny_general_2010,Ng_transpose,Beny_channel,Beny_perturb,Ng_unified,wangQuasiexactQuantumComputation2020}. 
Given a quantum code with encoding map $\mathcal{E}$
suffering from noise channel $\mathcal{N}(\rho) = \sum_u E_u \rho E_u^\dagger$, it satisfies:
\begin{equation}
    P E_u^\dagger E_v P = \lambda_{uv} P + PB_{uv}P.
    \label{eq:gKL}
\end{equation}
Here $P$ represents code subspace projection, $\lambda_{uv}$ is a constant satisfying $\lambda_{uv} = \lambda_{vu}^*$, $B_{uv}$ term captures deviation from exact QEC correction and can be viewed as logical error. If this logical error is sufficiently small, the recovery can be deemed high-precision, thereby enabling effective AQEC. In particular, an exact recovery channel $\mathcal{R}$ exists s.t. $\mathcal{R\circ N \circ E}= \mathrm{id}$, if and only if $B_{uv}=0$ \cite{knillTheoryQuantumErrorcorrecting1997}.


Using the framework of AQEC, we rigorously establish how the deviation from the exact Knill-Laflamme condition determines the optimal error threshold in general cases. 
This framework unifies the notion of error threshold for common QEC codes (for example stabilizer codes) as well as AQEC codes, which can only approximately preserve quantum information against local perturbations \cite{bravyiTopologicalQuantumOrder2010,BohdanowiczLDPC,BrandaoETH,Faistcovariant,yiComplexityOrderApproximate2023}. Remarkably, this method provides access to error thresholds of AQEC codes, which usually lack easily analyzable decoders.
We also examine two examples, the ordinary qudit stabilizer codes \cite{gottesmanFaultTolerantQuantumComputation1999} under stochastic noises where we exactly mapped the AQEC condition to maximum likelihood decoders and statistical mechanical models \cite{ref:dennis,ref:flammia}, and an imperfectly prepared toric code \cite{ref:guoyi,ref:fisher,zhao2023lattice} as an example of AQEC code that is unstable under local noises.

\section{AQEC relative entropy}
We conjecture that exploring the asymptotic behaviors of AQEC conditions with increasing code size $n$, is crucial for fully characterizing error thresholds.
To prove this statement, we examine a series of QEC codes, $\{\mathcal{C}_n\}_n^\infty$, and their respective noise channels, $\{\mathcal{N}_n\}_n^\infty$. The dimension of the code space, $K$, is dependent on the specific code; for instance, $K=4$ in qubit toric codes \cite{ref:dennis} and $\log K=\Theta(n)$ for good LDPC codes \cite{Panteleev,leverrier_tanner,Dinur}. In each instance of $\mathcal{C}_n$ and $\mathcal{N}_n$, Equation \eqref{eq:gKL} is fulfilled. Generally, $B_{uv}$ is nonzero for finite $n$. The hypothesis is that for systems below the threshold, $B_{uv}$ diminishes as $n\rightarrow \infty$, whereas it remains significant for systems above the threshold, even as $n\rightarrow \infty$.

The subtlety arises in quantifying the magnitude of $B_{uv}$. Within the chosen basis $\{\ket{q}\}_q$ of the code subspace, simply examining the matrix element $\bra{q_1} B_{uv} \ket{q_2}$ is insufficient to ascertain the threshold. In most cases $\bra{q_1} B_{uv} \ket{q_2}$ decays exponentially even above the threshold.  
Considering a surface code $ \ket{q}_{\text{SC}}, q=00,01,10,11$ affected by single-qubit Pauli $X$ errors $\mathcal{N} = \prod_i \mathcal{N}_i$, $ \mathcal{N}_i(\rho) = (1-p)\rho + pX \rho X$ \cite{ref:dennis}. It is easy to check that $\lambda_{uv}$ exhibits a scaling of $\mathcal{O}\left( (1-p)^{\frac{n}{2}} \right)$, whereas $\bra{q_1}_{\text{SC}} B_{uv} \ket{q_2}_{\text{SC}}$ demonstrates a scaling of $\mathcal{O}\left(p^\frac{\delta}{2} (1-p)^\frac{n-\delta}{2}\right)$, with $\delta \approx \sqrt{n}$. In the limit of $n\rightarrow \infty$, this is negligible for $p<1/2$. However, the actual threshold, identified using a maximum likelihood decoder (MLD), is approximately $p_c \sim 0.11$, indicating that matrix elements alone are inadequate for pinpointing the criticality in QEC systems. 
This phenomenon relates to the degeneracy of the code.
Therefore, we consider an entropic measure for a more precise threshold estimation. The parameters $\lambda_{uv}$ and operators $B_{uv}$ are reformulated in matrix form, pertinent to error configurations $uv$ and code words $q_1q_2$, as
\begin{equation}
    \Lambda_{uv,q_1q_2} = \frac{1}{K}\lambda_{uv} \delta_{q_1q_2},\quad  B_{uv,q_1q_2} = \frac{1}{K}\bra{q_1} B_{uv} \ket{q_2}.
\end{equation}
Here, $K$ represents the code space dimension. To ensure $\tr(\Lambda + B) = 1$, we introduce a factor of $\frac{1}{K}$, with both $\Lambda$ and $B$ being Hermitian \cite{beny_general_2010}. 
Furthermore, we can always assume $\tr_\mathcal{C}(B)=0$, where $\tr_\mathcal{C} (*)$ is the trace over code subspace, since the trace-nonzero part can always be 
absorbed into the definition of $\Lambda$. This is equivalent to choosing $\lambda_{uv} = \tr(PE^\dagger_u E_v)/K$. Note that both $\Lambda+B$ and $\Lambda$ are positive semi-definite. The \textit{AQEC relative entropy} is defined as
\begin{equation}
    S(\Lambda+B||\Lambda) = \tr\left\{(\Lambda+B)\left[\log(\Lambda+B)-\log\Lambda\right]\right\},
\end{equation}
which can be shown to satisfy (see Appendix \ref{sec:SI})
\begin{equation}
    0\leq S(\Lambda+B||\Lambda) \leq 2\log K.
    \label{eq:value}
\end{equation}
It measures the magnitude of $B$ relative to $\Lambda$, and intrinsically captures the logical error rate.
Notice that in a finite system, the lower bound in Eq. \eqref{eq:value} is saturated equality if and only if $B=0$ \cite{nielsenQuantumComputationQuantum2010}. In other words, the exact QEC condition is equivalent to $S(\Lambda+B||\Lambda) = 0$. As $S(\Lambda+B||\Lambda)$ is usually nonzero, the asymptotic behavior as $n\rightarrow \infty$ becomes crucial.


With the help of AQEC relative entropy, we define the \textit{intrinsic error threshold} as follows.
If the AQEC relative entropy vanishes in the large size limit $\lim_{n\rightarrow \infty} S(\Lambda+B||\Lambda) = 0$, we say that the QEC system is \textit{below the intrinsic error threshold}. Otherwise, the QEC system is \textit{above the intrinsic error threshold}. This intrinsic threshold demarcates different behaviors of AQEC relative entropy at large $n$, independent of decoder choice.

At the end of this section, we point out that there could be other quantities that measure the magnitude of $B$, for example using entanglement fidelity \cite{beny_general_2010}, operator norm \cite{Ng_transpose,Ng_unified} or other distance measures \cite{wangQuasiexactQuantumComputation2020}. 
We might alternatively extract the intrinsic error threshold from the asymptotic behavior of these measures since most AQEC measures can bound each other and become asymptotically equivalent in the large size limit. 
The central idea is that the large $n$ asymptotic behavior of AQEC measures determines the intrinsic error threshold.
We choose AQEC relative entropy since it is more tractable analytically, and the technical advantages are discussed in the following sections.


\section{Asymptotic recoverability}
Refs.~\cite{tyson_two-sided_2010,beny_general_2010} utilized worst-case entanglement fidelity to measure the deviation of AQEC from QEC condition, limited to a fixed code size. Our work examines the implication of AQEC relative entropy on recovery channels in the asymptotic large code size limit.

We use entanglement fidelity~\cite{schumacher_quantum_1996,schumacherSendingEntanglementNoisy1996,barnumInformationTransmissionNoisy1998} to quantify the success of recovery,
\begin{equation}
\begin{aligned}
    &F_e (\rho_0,\mathcal{R\circ N \circ E}) = \\
    &\bra{\Psi_{AQ}} (\mathrm{id}_A \otimes \mathcal{R\circ N \circ E})(\ket{\Psi_{AQ}}\bra{\Psi_{AQ}}) \ket{\Psi_{AQ}},
\end{aligned}
\end{equation}
where $\ket{\Psi_{AQ}}$ is the purification of $\rho_0$ by introducing ancilla system $A$.
In our case, the initial state $\rho_0$ is chosen as the maximally mixed logical state $ \rho_0=I_K/K $, making $F_e (I_K/K,\mathcal{R\circ N \circ E}) $ indicative of the QEC success in an average sense. 
Indeed, we can define the average recovery fidelity as
\begin{equation}
    \overline{F}(\mathcal{R\circ N \circ E}) = \int d\psi \bra{\psi}\mathcal{R\circ N \circ E}(\ket{\psi}\bra{\psi}) \ket{\psi},
\end{equation}
where $\psi$ belongs to the $K$ dimensional Hilbert space and the integration is over the Fubini-Study measure. It is related to the entanglement fidelity by \cite{horodeckiGeneralTeleportationChannel1999,woodTensorNetworksGraphical2015}
\begin{equation}
    \overline{F}(\mathcal{R\circ N \circ E}) = \frac{1+K F_e (I_K/K,\mathcal{R\circ N \circ E})}{1+K}.
\end{equation}
From now on we abbreviate $ F_e (\mathcal{R\circ N \circ E})= F_e (I_K/K,\mathcal{R\circ N \circ E})$.
Note that $F_e (\mathcal{R\circ N \circ E})=$ = 1 if and only if $\mathcal{R\circ N \circ E}= \mathrm{id}$. In general, it is smaller than $1$ in the finite-size case. So considering the limit $ \lim_{n\rightarrow \infty} F_e (\mathcal{R\circ N \circ E}) $ helps determine if noise is \textit{asymptotically recovered} from the noise. If $ \lim_{n\rightarrow \infty} F_e (\mathcal{R\circ N \circ E}) = 1$, it suggests that the system approaches exact recovery as we increase its size, namely asymptotically recoverable. It is related to the AQEC relative entropy as follows.

\begin{theorem}
\label{thm1}
Given a family of $\{\mathcal{C}_n\}_n^\infty$ with noise channels $\{\mathcal{N}_n\}_n^\infty$, consider the large size limit $n\rightarrow \infty$,
\begin{enumerate}[(1)]
    \item Below the intrinsic error threshold, i.e. $\lim_{n\rightarrow \infty} S(\Lambda+B||\Lambda) = 0$, there \textbf{exists} a family of recovery map $\{\mathcal{R}_n\}_n$, such that the entanglement fidelity of the whole QEC process satisfies
\begin{equation}
    \lim_{n\rightarrow \infty} F_e (\mathcal{R\circ N \circ E})=1.
    \label{eq:fidelity}
\end{equation}
\item Above the intrinsic error threshold, i.e. $S(\Lambda+B||\Lambda)$ does not converge to $0$, if $K=\mathcal{O}(1)$, then the entanglement fidelity $F_e (\mathcal{R\circ N \circ E})$ cannot converges to $1$ for an \textbf{arbitrary} family of recovery map $\{\mathcal{R}_n\}_n$.
\item Let $K$ be a parameter that diverges with $n$, such that $K=\omega(1)$. If $s(\Lambda+B||\Lambda) \equiv S(\Lambda+B||\Lambda)/\log K$ does not converge to $0$, the entanglement fidelity $F_e (\mathcal{R\circ N \circ E})$ cannot converges to $1$ for an \textbf{arbitrary} family of recovery map $\{\mathcal{R}_n\}_n$.
\end{enumerate}
\end{theorem}

The proof of the theorem utilizes the following lemma concerning another quantity called coherent information \cite{schumacher_quantum_1996,schumacherSendingEntanglementNoisy1996,barnumInformationTransmissionNoisy1998,schumacher_approximate_2001}
\begin{equation}
    I_c(\rho_{0},\mathcal{ N \circ E}) = S(\rho_Q)-S(\rho_{AQ}),
    \label{eq:coherent}
\end{equation}
where $\rho_{AQ} = id\otimes \mathcal{N} (\ket{\Psi_{AQ}} \bra{\Psi_{AQ}})$ with the noise $\mathcal{N}$ acting on subsystem $Q$, and $\rho_Q = \tr_{A} \rho_{AQ}$. The coherent information assesses the preservation of information under a noisy process.
\begin{lemma}
\label{lemma1}
For a QEC code $\mathcal{C}$ under a noise channel $\mathcal{N}$, the AQEC relative entropy is related to the coherent information through:
\begin{equation}
    S(\Lambda+B||\Lambda) = -I_c(I/K,\mathcal{ N \circ E})+ \log K.
\end{equation}
\end{lemma}
\begin{remark}
The lemma relates the AQEC relative entropy defined from the generalized Knill-Laflamme condition Eq. \eqref{eq:gKL} to the coherent information. The coherent information examines the effect of noise channels on the code subsystem while the AQEC relative entropy studies the error subsystem. Following the spirit of Ref. \cite{beny_general_2010}, AQEC relative entropy is the complementary form of coherent information and is more convenient in certain calculations. We again have chosen $\rho_0 = I/K$ to yield an average quantity, and it is related to the choice of $\lambda_{uv} = \tr(PE^\dagger_u E_v)/K$.
The proof of the Lemma can be found in Appendix \ref{sec:SI}. 
\end{remark}

\begin{proof}[Proof of Theorem \ref{thm1}]

Given coherent information $I_c(I/K,\mathcal{ N \circ E})$, there exist a recovery channel $R$ such that 
\cite{schumacher_approximate_2001,buscemiEntanglementMeasuresApproximate2008}
\begin{equation}
    1\geq F_e (\mathcal{R\circ N \circ E}) \geq 1-\sqrt{2(-I_c(I/K,\mathcal{ N \circ E})+\log K)},
\end{equation}
and for an arbitrary quantum channel $\mathcal{R}$ we have \cite{barnumInformationTransmissionNoisy1998,buscemiEntanglementMeasuresApproximate2008}
\begin{equation}
    0 \leq -I_c(I/K,\mathcal{ N \circ E})+\log K \leq 2f(1-F_e (\mathcal{R\circ N \circ E})),
\end{equation}
where $f(x)=h(x) + x \log(K^2-1)$, and $ h(x) = -x\log x - (1-x) \log(1-x)$ is the binary entropy.
Now for a family of QEC code labeled by system size $n$ and has a limit $n\rightarrow \infty$, the above two statements hold for all $n$, that is there exists a family of recovery channels $\{\mathcal{R}_n\}_n$ such that
\begin{equation}
    0\leq r(\mathcal{R}_n \circ \mathcal{N}_n \circ \mathcal{E}_n) \leq \sqrt{2S(\Lambda_n+B_n||\Lambda_n)},
    \label{eq:below}
\end{equation}
while for every possible family of recovery channels $\{\mathcal{R}_n\}_n$,
\begin{equation}
    0 \leq S(\Lambda_n+B_n||\Lambda_n) \leq 2f(r (\mathcal{R}_n \circ \mathcal{N}_n \circ \mathcal{E}_n) ),
    \label{eq:above}
\end{equation}
where we have used Lemma \ref{lemma1}.
Here the subscripts label the system size, and we define infidelity $r(\mathcal{R}_n \circ \mathcal{N}_n \circ \mathcal{E}_n) = 1-F_e(\mathcal{R}_n \circ \mathcal{N}_n \circ \mathcal{E}_n)$.
Below the intrinsic threshold $\lim_{n\rightarrow \infty} S(\Lambda_n+B_n||\Lambda_n) = 0$, Eq. \eqref{eq:below} tells us $\lim_{n\rightarrow \infty} r(\mathcal{R}_n \circ \mathcal{N}_n \circ \mathcal{E}_n) = 0$. Above the threshold where $\lim_{n\rightarrow \infty} S(\Lambda_n+B_n||\Lambda_n) > 0$ or diverges, 
$\lim_{n\rightarrow \infty} r(\mathcal{R}_n \circ \mathcal{N}_n \circ \mathcal{E}_n)$ must also $>0$ or diverges for the finite $K$ case, otherwise Eq. \eqref{eq:above} leads to contradiction. 
For the case $K_n$ depend on $n$ and $K_n \rightarrow \infty$ as $n\rightarrow \infty$. if the infidelity $r(\mathcal{R}_n \circ \mathcal{N}_n \circ \mathcal{E}_n) \rightarrow 1$, we  conclude that the density of AQEC relative entropy $s(\Lambda+B||\Lambda) = S(\Lambda+B||\Lambda)/\log K_n$ converges to zero rather than $S(\Lambda+B||\Lambda)$ itself.
    
\end{proof}

The theorem specifies how the AQEC relative entropy in the large size limit determines the asymptotic recoverability below or above the intrinsic threshold.
Notably, with $ K $ diverging with $n$, Eq.\eqref{eq:above} above implies $ 1-F_e (\mathcal{R\circ N \circ E})=\Omega(1/\log K) $, i.e. the infidelity decays no faster than $1/\log K$ above threshold, due to $ f(x) $'s $ K $-dependency.
Instead, only if $ S(\Lambda+B||\Lambda) \rightarrow 2c \log K$ and $0<c\leq 1$ is a finite constant can we get a lower bounded infidelity $r (\mathcal{R} \circ \mathcal{N} \circ \mathcal{E}) \geq c/2 > 0$.
Therefore, the ``logical qubit number'' $ k \propto \log K $ can be used as the denominator to evaluate the \textit{ density of AQEC relative entropy}, $s(\Lambda+B||\Lambda)=S(\Lambda+B||\Lambda)/\log K$, aiding in assessing the failure of asymptotic recoverability.
Intuitively, the entropy measure $S(\Lambda+B||\Lambda)$ is more sensitive to errors than the fidelity measure $F_e(\mathcal{R\circ N\circ E})$, since it captures smaller physical noise in its nonzero-ness, while its convergence to zero guarantees asymptotic recoverability.
Notice that we use the quantity $K$, the code space dimension, rather than the logical qubit or qudit number $k$, since in the most general cases $K$ could be an arbitrary integer, and the logical qudit number is not always well-defined. In the next section, we will see that for a qudit stabilizer code in prime local dimension $d$, $K$ is always factorized into $k^d$, so the standard notation of a $[[n,k,d]]$ code is then valid. 

Till now, we have not specified a decoder, which is essentially a specific choice of recovery channel, denoted as $\mathcal{R}_{\text{de}}$. In scenarios with finite $K$, achieving $\lim_{n\rightarrow \infty} F_e =1$ is impossible when above the intrinsic threshold. Below this threshold, although some recovery channel $\mathcal{R}$ may achieve perfect entanglement fidelity, $\mathcal{R}_{\text{de}}$ may not be equally effective. Thus, the intrinsic error threshold sets an upper bound on decoding thresholds, i.e. the optimal threshold. With $K$ diverging, the density of AQEC relative entropy $s(\Lambda+B||\Lambda)$ serves as a metric to upper bound decoding thresholds.

The two phases of the QEC system have just been discussed. However, the critical examination of an intrinsic threshold, or a phase transition, remains. Subsequent sections address this through specific examples using our framework.
{In general two mechanisms could result in the extra $B$ term in Eq. \eqref{eq:gKL}. One is that a realistic noise channel $\mathcal{N}$ normally contains nonlocal Kraus operators $E_{u}$. For the ordinary single-qubit stochastic error models, chances are that certain $E^\dagger_u E_v$ operators are supported on a region larger than the code distance, which leads to nonzero $B_{uv}$. Another one is that the code space itself is imperfect. In this case, even if the error operator $E^\dagger_u E_v$ is supported on a local region, $B_{uv}$ can also acquire a small but nonzero value. Such codes are known as AQEC codes. For example, suppose that a many-body system is in a topologically ordered phase, but there is a perturbation in Hamiltonian such that it does not stay at the stable fix point, then the local error operator will lead to an exponentially small extra term $B_{uv,q_1q_2} \sim \exp(- \mathcal{O}(n))$ \cite{bravyiTopologicalQuantumOrder2010}. Another example is when the preparation circuit of a stabilizer code suffers from coherent noises \cite{ref:guoyi,ref:fisher,zhao2023lattice}, which we mainly focus on.  Our following examples cover both two mechanisms. In Sec. \ref{sec:stabilizer} we study perfect stabilizer codes under stochastic noises, while in Sec. \ref{sec:weak} we study the imperfectly prepared toric code. Our discussion shows how the asymptotic behavior of AQEC relative entropy determines the error threshold in both cases, thus providing a unified framework of error threshold for both exact and approximate QEC codes.  }



\section{Stabilizer codes}
\label{sec:stabilizer}
\subsection{phase space formalism and definition of stabilizer codes}
We first consider qudit stabilizer codes \cite{gottesman1997stabilizer,gottesmanFaultTolerantQuantumComputation1999,gross} within the phase space formalism.
For a $n$ qudit system with local Hilbert space dimension $d\geq 2$, the Heisenberg-Weyl operator is defined as:
\begin{equation}
    T(v) = T(v_p,v_q) = \omega^{-\frac{1}{2} v_p^T v_q} Z^{v_p} X^{v_q}.
\end{equation}
Here $\omega = e^{i2\pi/d}$, both $v_p$ and $v_q$ are $n$ dimensional $\mathbb{Z}_d$ valued vectors, and $v=(v_p,v_q) \in \mathbb{Z}_d^{2n}=\mathcal{V}$. We restrict to prime dimension $d$ for convenience, and here all arithmetic is done modulo $d$. $X$ and $Z$ are generalized Pauli operators,
\begin{equation}
    X\ket{q} = \ket{q+1} , \quad Z\ket{q} = \omega^q \ket{q}, \quad q \in \mathbb{Z}_d.
\end{equation}
$Z^{v_p}$ and $X^{v_q}$ denotes the Weyl strings acting on $n$ qudits
\begin{equation}
    Z^{v_p} = \bigotimes_{i=1}^n Z^{v_{pi}}, \quad X^{v_q} = \bigotimes_{i=1}^n X^{v_{qi}}.
\end{equation}
All such $T(v)$ generates the discrete Heisenberg-Weyl group.
In general, $v$ can be interpreted as a point in the classical phase space $\mathcal{V}=\mathbb{Z}_d^{2n}$. Specifically, $v_q$ is the coordinate vector and $v_p$ is the momentum vector. The action of $T(v)$ leads to a translation in both the coordinate and momentum space, so it is also called the "Displacement operator".
The basic algebraic relation of Weyl operators is
\begin{equation}
    T(v) T(u) = \omega^{\frac{1}{2} [v,u]} T(v+u).
    \label{eq:algebra}
\end{equation}
Here $[v,u] = v^T \Omega u$ is the symplectic inner product on $\mathcal{V}$, where
\begin{equation}
    \Omega = \begin{pmatrix} 
    0_{n\times n} & I_{n\times n}\\
    -I_{n\times n} & 0_{n\times n}\\
    \end{pmatrix}.
\end{equation}
We can see that $T$ defines a projective representation of $\mathcal{V}$ on the Hilbert space. 
Note that we slightly abuse the square bracket notation to represent both the symplectic inner product and operator commutator. Without leading to confusion, it stands for symplectic inner product when acting on $\mathbb{Z}_d$ vectors and for operator commutator when acting on quantum operators. The reason of doing so is to indicate that they are related, since the symplectic inner product completely encodes the information of the commutator of Heisenberg-Weyl operators. 
Besides, for odd prime $d$, the notation $\frac{1}{2}$ here actually stands for the multiplicative inverse of $2$ modulo $d$, i.e. $2^{-1} = (d+1)/2$, which is an element of $\mathbb{Z}_d$. For the $d=2$ qubit case, we should interpret it as $\omega^{1/2} = i$.

Within the phase space formalism, let us define stabilizer codes. Given a subspace $\mathcal{M}$ of $\mathcal{V}$, $\mathcal{M}$ is called isotropic if and only if
\begin{equation}
    [m_1,m_2]=0, \quad \forall m_1,m_2\in \mathcal{M}.
\end{equation}
If $\mathcal{M}$ is isotropic, then all corresponding Weyl operators commute with each other, $[T(m_1), T(m_2)]=0$. In that case, $T$ is an isomorphism between $\mathcal{M}$ and an abelian subgroup of the discrete Heisenberg-Weyl group, which is called stabilizer group. Actually, the elements in stabilizer group can be redefined with some phase factor, but we will not keep track of the phases here since they are irrelevant in our discussion. 

Given an isotropic subspace $\mathcal{M}$, the corresponding stabilizer group is defined as $T(\mathcal{M})$, which is the image of mapping $\mathcal{M}$ onto the operator space. The associated stabilizer code subspace $\mathcal{C}$ is defined as the maximal subspace of the Hilbert space which satisfies
\begin{equation}
    T(m) \ket{\psi} = \ket{\psi}, \quad \forall m \in \mathcal{M}, \quad \forall \ket{\psi} \in \mathcal{C}.
\end{equation}
The cardinal of $\mathcal{M}$ has to be less than $d^n$ such that the eigenspace $\mathcal{C}$ has degeneracy. In fact, the dimension of $\mathcal{C}$ (denoted as $K$) must satisfy $K=d^n/|\mathcal{M}|$ \cite{gheorghiu_standard_2014}.
We specify $\mathcal{M}$ by giving a basis $\{m_1,\cdots,m_r\}$. This basis is mapped to a set of generators $\{T(m_1),\cdots,T(m_r)\}$ of stabilizer group by $T$.

Here a remark should be made about the local dimension $d$. If $d$ is a prime number $d=2,3,5,\cdots$, then $\mathbb{Z}_d$ is a field and $\mathcal{V}$ is a true vector space. Thus most of the results in the qubit case can be naturally generalized to prime dimensions.
For example, the dimension of the code subspace will be $K=d^{n-r}=d^k$, which means that there are $k$ logical qudits, leading to an $[[n,k]]$ code.
But if $d$ is non-prime, $\mathcal{V}$ is mathematically a $\mathbb{Z_d}$ module, and subtleties arise in the algebraic structure \cite{gross,gheorghiu_standard_2014}. In that case, $K$ generally cannot be written in the form $d^k$.

Now we want to find the logical operators within the Heisenberg-Weyl group. Notice that any logical operator must commute with all the stabilizers so that they do not affect the error syndrome. We define the symplectic complement of $M$ as \cite{gross}
\begin{equation}
    \mathcal{M}^\perp =\{ v\in \mathcal{V} | [m,v]=0, \quad \forall m \in \mathcal{M} \}.
    \label{eq:Mperp}
\end{equation}
The suitable logical Weyl operators must be in $\mathcal{M}^\perp$. Since the operators in $\mathcal{M}$ act trivially on the code subspace, the space of logical Weyl operators will be chosen as $\mathcal{L} = \mathcal{M}^\perp/\mathcal{M}$. For each equivalent class $[l] \in \mathcal{L}$, we choose a representative element $l$ and define the corresponding logical operator as $[l] \mapsto T(l)$. Since the size of $\mathcal{M}^\perp$ satisfies $|\mathcal{M}^\perp| = d^{2n}/|\mathcal{M}| = d^{n+k}$ \cite{gross}, we have $|\mathcal{L}| = K^2$. The code distance $\delta$ is defined as the minimal weight of a nontrivial logical operator $T(l), l\in \mathcal{M}^\perp-\mathcal{M}$, while 'weight' means the number of data qudits that $T(l)$ nontrivially acts on.

From now on we consider only $[[n,k,\delta]]$ code in prime local dimensions $d$ for simplicity. It follows that $\mathcal{V}$ is a $2n$ dimensional $\mathbb{Z}_d$ vector space, and $\mathcal{M}$, $\mathcal{M}^\perp$ are respectively its $r=n-k$ and $n+k$ dimensional subspace. $\mathcal{L}$ is a quotient vector space with dimension $2k$. We abbreviate the logical class $[l]$ as its representative element $l$ without leading to misunderstanding.

\subsection{threshold of maximum likelihood decoder}
We assume that the data qudits suffer from stochastic weyl errors, 
\begin{equation}
    \mathcal{N}(\rho_0)=\sum_{\eta\in \mathcal{V}} \Pr(\eta) T(\eta) \rho_0 T(\eta)^\dagger.
    \label{eq:error}
\end{equation}
MLD~\cite{ref:dennis,ref:flammia}, the optimal decoder for this QEC system, selects recovery operators by assessing the combined likelihood of errors that produce the same syndrome. Notice that each error configuration $\eta \in \mathcal{V}$  can be decomposed as $\eta = s + l + m$, where $s \in \mathcal{S}=\mathcal{V}/\mathcal{M}^\perp$ stands for a particular syndrome, $l\in \mathcal{L}$ denotes logical classes and $m \in \mathcal{M}$.
Note that $\mathcal{V}/\mathcal{M}^\perp \cong (\mathcal{V}/\mathcal{M})/\mathcal{L}$ and $\dim \mathcal{S} = n-k$. Since $\dim \mathcal{S}  + \dim\mathcal{L} + \dim \mathcal{M} = \dim \mathcal{V}$, the decomposition is unique. However, the choice of the representative element of the classes $s\in \mathcal{S}$ and $l \in \mathcal{L}$ can be arbitrary.
Each sydrome $s$ contains $d^k$ logical classes $l$ while each logical class contains $d^{n-k}$ stabilizers $m$.  Since stabilizers act trivially on code subspace, we are only concerned about the error equivalent classes when deciding the recovery operator. 
An error equivalent class is specified by syndrome $s$ and logical class $l$, and its joint probability is
\begin{equation}
    \Pr(s,l) = \sum_{m\in \mathcal{M}} \Pr(s+l+m).
    \label{eq:joint}
\end{equation}
We denote the corresponding random variables as $S$ and $L$. Given a syndrome $s$, the maximum likelihood decoder (MLD) chooses recovery operator $T(s+l)$ with the largest conditional probability
\begin{equation}
    \Pr(l|s) = \Pr(s,l)/\Pr(s).
    \label{eq:condition}
\end{equation}
Without losing generality, we redefine the representative configuration of syndrome $s$ such that $\Pr(l=0|s)$ maximizes the likelihood.
The success rate of MLD is commonly measured by the probability of $l=0$ logical class,
\begin{equation}
    \Pr(l) = \sum_{s\in\mathcal{S}}\Pr(s,l),
\end{equation}
\begin{equation}
    \Pr(\text{MLD success}) = \Pr(l=0).
\end{equation}
We also introduce another measure of success, which is the Shannon entropy of $L$ conditioned on $S$,
\begin{equation}
\begin{aligned}
    H(L|S) &= -\sum_{s\in\mathcal{S}} \sum_{l\in\mathcal{L}} \Pr(s,l) \log \Pr(l|s)\\& = \sum_{s\in\mathcal{S}}\Pr(s)H(L|S=s),    
\end{aligned}
\label{eq:shannon}
\end{equation}
where $H(L|S=s)$ is the Shannon entropy when syndrome $s$ is specified,
\begin{equation}
    H(L|S=s) = -  \sum_{l\in\mathcal{L}} \Pr(l|s) \log \Pr(l|s).
\end{equation}
$H(L|S)$ measures the uncertainty in MLD when deducing the logical class $l$ from the syndrome $s$.

A standard approach to the MLD threshold problem is mapping the error class probability to a statistical mechanical (SM) partition function $\Pr(s,l)= Z(\eta)$ with quenched disorder $\eta$ \cite{ref:flammia}, generalizing previous SM mapping constructions \cite{ref:dennis,colorcode,Bombin-arxiv,kubica_three-dimensional_2018,kovalevNumericalAnalyticalBounds2018}. 
For stabilizer codes and Weyl errors, the SM mapping always exists thanks to the structure of Clifford group. According to the Gottesman-Knill theorem \cite{nielsenQuantumComputationQuantum2010,gross},  The stabilizer code words and Weyl noise operators both acquire phase space descriptions, thus the noise probability distributions on the classical phase space can be mapped to the Boltzmann weights of the classical SM model.

Following this spirit, we can write the probability of error equivalent classes $[\eta] \in \mathcal{V}/\mathcal{M}$ into classical partition functions,
\begin{equation}
\begin{aligned}
    &  Z(\eta) = \Pr([\eta]) = \Pr(s,l) =\\&   \sum_{m \in \mathcal{M}}  \Pr(\eta + m)= \sum_{c\in\mathbb{Z}_d^r} \exp\left(-H_\eta(c) \right),\\
    &H_\eta(c) = - \log\Pr(\eta + Mc), \quad c\in\mathbb{Z}_d^r,
\end{aligned}
\label{eq:partition}
\end{equation}
were $M=(m_1,\cdots,m_r)$ is the $n\times r$ matrix where each column is a basis vector of $\mathcal{M}$, $c\in\mathbb{Z}_d^r$ serves as the SM d.o.f. and $\eta$ inherits a quenched disorder configuration from error probability $\Pr(\eta)$. For example, for single-qudit noise channels 
$\mathcal{N}=\otimes_{i=1}^n \mathcal{N}_i$ where $\mathcal{N}_i(\rho_0) = \sum_{\eta_i \in \mathbb{Z}_d^2} p_i(\eta_i) T(\eta_i) \rho_0 T(\eta_i)^\dagger$, the Hamiltonian $H_\eta$ has a more explicit form
\begin{equation}
\begin{aligned}
    &H_\eta(c) = - \sum_{i=1}^n \sum_{v_i\in\mathbb{Z}_d^2} J_i(v_i) \omega^{[v_i,\eta_i + M_i c]},\\
    &J_i(v_i) = \frac{1}{d^2}\sum_{u_i \in \mathbb{Z}_d^2}  \omega^{-[v_i,u_i]} \log p_i(u_i),
\end{aligned}
\end{equation}
Where $M_i$ is the $i$-th row of $M$. Or suppose that the noises are locally correlated and factored into the form
\begin{equation}
    \Pr(\eta) = \prod_{R} p_{R} (\eta_R),
\end{equation}
where $R$ labels possibly overlapped regions on the lattice and $\eta_R$ is the restriction of $\eta$ on the subsystem $R$.
In this case the SM model becomes
\begin{equation}
\begin{aligned}
    &H_\eta(c) = - \sum_{R} \sum_{v_R\in\mathbb{Z}_d^2} J_R(v_R) \omega^{[v_R,\eta_R + M_R c]},\\
    &J_R(v_R) = \frac{1}{d^{2|R|}}\sum_{u_R \in \mathbb{Z}_d^2}  \omega^{-[v_R,u_R]} \log p_R(u_R),
\end{aligned}
\end{equation}
where $M_R$ is now a $|R| \times r$ matrix constituted by the $i\in R$ rows of $M$.
Suppose that the error rate of $\mathcal{N}$ can be tuned, then the error threshold of MLD is identified with the order-disorder phase transition of the SM model. The order parameter of such a phase transition is traditionally chosen as the average free energy cost of a nontrivial logical operator $l \neq 0$,
\begin{equation}
    \begin{aligned}
        &\overline{\Delta_l} = \sum_{\eta \in \mathcal{V}} \Pr(\eta)\Delta(\eta,l),\quad \Delta(\eta,l) = - \log \frac{Z(\eta + l)}{Z(\eta)},
    \end{aligned}
\end{equation}
with $\Delta(\eta,l)$ the free energy cost under a particular disorder configuration $\eta$. It diverges below the decoding threshold and in most cases finite above the decoding threshold. Notice that for $l=0$, $\Delta(\eta,l)$ is trivially $0$.

\subsection{main results on stabilizer codes}
A key result of ours for stabilizer codes is the following.
\begin{lemma}
\label{lemma2}
    For a qudit stabilizer code in prime $d$ and a stochastic Weyl noise channel, the AQEC relative entropy, probability of logical classes and order parameter of the SM model are related by
    \begin{equation}
    \begin{aligned}
        S(\Lambda+B||\Lambda) &=H(L|S)\\&= \sum_{\eta \in \mathcal{V}} \Pr(\eta) \log \left\{ \sum_{l \in \mathcal{L}} \exp\left[ -\Delta(\eta,l)\right]\right\},
    \end{aligned}
    \label{eq:main}
    \end{equation}
    where $H(L|S)$ is the Shannon conditional entropy of logical class $l$ given syndrome $s$, and the last quantity is a generalized version of the homological difference  \cite{kovalevNumericalAnalyticalBounds2018}.
\end{lemma}

\begin{remark}
We applied the replica method in the derivation of Eq. \eqref{eq:main}. The rigorousness is guaranteed by Carlson's theorem \cite{boas_entire_1973,wittenOpenStringsRindler2019,dhokerAlternativeMethodExtracting2021}. We calculate the AQEC relative entropy through the limit
\begin{equation}
    S(\Lambda+B||\Lambda) = \lim_{R\rightarrow 1}\frac{1}{R-1} \log \frac{\tr((\Lambda+B)^R)}{\tr(\Lambda^{R})}.
\end{equation}
The r.h.s. is evaluated by fixing $R$ as integers $R>1$,
\begin{equation}
    \begin{aligned}
        &\tr(\Lambda^R)  = \frac{1}{d^{k(R-1)}} \sum_{\eta \in \mathcal{V}} \Pr(\eta )Z(\eta)^{R-1},\\
        &\tr((\Lambda+B)^R)  = \frac{1}{d^{k(R-1)}} \sum_{\eta \in \mathcal{V}} \Pr(\eta )\left[\sum_{l\in\mathcal{L}}Z(\eta+l)\right]^{R-1},
    \end{aligned}
    \label{eq:partition1}
\end{equation}
and then taking the $R\rightarrow 1$ limit after analytical continuation. A detailed proof is in Appendix \ref{sec:SII}. Note that the generalized homological difference serves as a better-behaved order parameter compared to the simple disorder averaged $\overline{\Delta_l}$. In particular, $\overline{\Delta_l}$ could possibly diverge even above the threshold when a single unlikely syndrome can be perfectly decoded and all others provide a logical error, and thus fail to determine the threshold position \cite{ref:flammia}. This problem is completely fixed by using AQEC relative entropy $S(\Lambda+B||\Lambda)$ or Shannon conditional entropy $H(L|S)$ since they are always nonzero above the threshold, even in the above extreme case.
\end{remark}
Notice that the r.h.s. of Eq. \eqref{eq:partition1} are the replica partition functions \cite{nishimoriStatisticalPhysicsSpin2001} of the quenched disordered SM model $Z(\eta)$ (inserted with domain wall configurations $l$ in the second line).
 In classical spin glass theories, in order to compute the disorder averaged free energy $F=-\sum_{\eta} \Pr(\eta) \log Z(\eta)$ in an accessible way, one often averages the $R$-th power of partition function first and then take the $R\rightarrow 1$ limit after analytical continuation,
\begin{equation}
    F = \lim_{R\rightarrow 1} \frac{1-\sum_{\eta} \Pr(\eta)Z(\eta)^{R-1}}{R-1}.
\end{equation}
The replica partition function is defined as $Z^{(R-1)}= \sum_{\eta} \Pr(\eta)Z(\eta)^{R-1}$.
It suggests that the SM mapping can be derived from the intrinsic properties of the code and noise, following the spirit of Fan et al.'s example \cite{ashvin}. 
Note that we considered a different SM mapping to avoid the potentially complex Boltzmann weights in their original SM models when applied to $d\geq 3$ qudit systems.
For stabilizer codes and Weyl noises, the SM mapping should always valid which is guaranteed by the Gottesman-Knill theorem \cite{nielsenQuantumComputationQuantum2010,gross}, since they belong to Clifford algebra and are intrinsically classical. From an algebraic point of view, the relevant degree of freedom can be  mapped to classical ones in the phase space.
Therefore, $\tr(\Lambda^R)$ (or $Z(\eta)$) is a more suitable choice for the partition function to circumvent complex weights. Additionally, it enables us to directly take the replica limit $R\rightarrow 1$ on the SM model side, which was not achieved in Ref. \cite{ashvin}.



Eq.\eqref{eq:main} suggests a link between the intrinsic and MLD thresholds. Given that the MLD threshold is typically evaluated by the asymptotic behavior of the success probability $\Pr(l=0) = 1$ \cite{ref:flammia,kovalevNumericalAnalyticalBounds2018}, we can establish a corresponding relation.
\begin{theorem}
\label{thm2}
    For a family of qudit stabilizer codes in prime $d$ and stochastic Weyl noise channels, in the large size limit $n\rightarrow \infty$,
    \begin{enumerate}[(1)]
    \item below the intrinsic error threshold, i.e. $\lim_{n\rightarrow \infty} S(\Lambda+B||\Lambda) = 0$, the success probability of MLD converges to $1$,
    \begin{equation}
        \lim_{n\rightarrow \infty} \Pr(l=0) = 1.
    \end{equation}
    \item if the logical qudit number is finite $k=\mathcal{O}(1)$, then $\lim_{n\rightarrow \infty}\Pr(l=0) = 1$ implies the QEC system is below the intrinsic error threshold $\lim_{n\rightarrow \infty} S(\Lambda+B||\Lambda) = 0$.
    \item if $k$ diverges with $n$, $k=\omega(1)$, then $\lim_{n\rightarrow \infty}\Pr(l=0) = 1$ implies the density of AQEC relative entropy $s(\Lambda+B||\Lambda) \equiv S(\Lambda+B||\Lambda)/k$ converges to $0$.
    \end{enumerate}
\end{theorem}

\begin{proof}[Proof of Theorem \ref{thm2}]
Assume $\lim_{n\rightarrow \infty}S(\Lambda+B||\Lambda) = 0$, the first proposition follows from the concavity of logarithmic function,
\begin{equation}
    \begin{aligned}
        &1\geq \Pr(l=0) = \sum_{s\in\mathcal{S}} \Pr(s) \Pr(l=0|s) \\&\geq \exp \left[\sum_{s\in\mathcal{S}} \Pr(s) \log \Pr(l=0|s)\right] \\&= \exp \left[\sum_{s\in\mathcal{S}} \sum_{l\in\mathcal{L}} \Pr(s,l) \log \Pr(0|s)\right] \\&\geq \exp \left[\sum_{s\in\mathcal{S}} \sum_{l\in\mathcal{L}} \Pr(s,l) \log \Pr(l|s)\right] = \exp \left[-H(L|S)\right]. 
    \end{aligned}
\end{equation}
Note that we have used $\Pr(0|s) \geq \Pr(l|s)$, $\forall l \in \mathcal{L}$. So $\lim_{n\rightarrow \infty} \Pr(l=0) = 1$. 

The second and third propositions follow that conditioning reduces entropy,
\begin{equation}
    0 \leq H(L|S) \leq H(L),
\end{equation}
where $H(L) = -\sum_{l\in \mathcal{L}} \Pr(l) \log \Pr(l)$. 
Fixing the success probability $\Pr(l=0)$, the Shannon entropy $H(L)$ is maximized when the remaining logical classes are uniformly distributed,
\begin{equation}
\begin{aligned}
    &0 \leq H(L|S) \leq H(L) \\&\leq - \Pr(l=0) \log \Pr(l=0)\\& - (K^2-1) \frac{1- \Pr(l=0)}{K^2-1} \log\frac{1- \Pr(l=0)}{K^2-1}\\&= f(1- \Pr(l=0)).
\end{aligned}
\label{eq:succ-prob}
\end{equation}
The above inequality is called the classical Fano's inequality \cite{coverElementsInformationTheory2006}.
For finite $k = \log K / \log d$, $\Pr(l=0)\rightarrow 1$ implies $\lim_{n\rightarrow \infty}S(\Lambda+B||\Lambda) = \lim_{n\rightarrow \infty} H(L|S) = 0$. For $k \rightarrow \infty$, $\Pr(l=0)\rightarrow 1$ in turn implies $\lim_{n\rightarrow \infty}s(\Lambda+B||\Lambda)  = 0$.
    
\end{proof}

The first two propositions in Thm. \ref{thm2} suggests that the intrinsic threshold is exactly the same as the MLD threshold for those codes with finite many logical qudits, if the decodable region is defined by $\lim_{n\rightarrow \infty}\Pr(l=0) = 1$. For $k\rightarrow \infty$, we similarly can only conclude $1- \Pr(l= 0) = \Omega(1/k)$ when above the intrinsic threshold.
The theorem demonstrates that applying our decoder-independent method to stabilizer codes with Weyl noise yields a threshold equivalent to that obtained from the maximum likelihood decoder used in other methods. The optimality of the maximum likelihood decoding threshold is a direct corollary of Theorem \ref{thm1} and \ref{thm2}. 
Also, the intrinsic threshold of QEC systems, as indicated by the second equality of Eq.\eqref{eq:main}, is contingent on the phase transition of the corresponding SM model. Notable instances encompass qubit or qudit toric code \cite{ref:dennis,ashvin}, color code \cite{colorcode,Bombin-arxiv,kubica_three-dimensional_2018}, qubit or qudit hypergraph-product code \cite{kovalevNumericalAnalyticalBounds2018,jiangDualityFreeEnergy2019,Jiang}, Hyperbolic surface code \cite{breuckmannConstructionsNoiseThreshold2016,jiangDualityFreeEnergy2019,Jiang} and X-Cube code \cite{song_optimal_2022} under single-qubit Pauli noise,
toric code under locally correlated Pauli noise \cite{ref:flammia}. 

For example, let us assume geometric locality for both stabilizer generators and the error channel. In the ordered phase, nontrivial logical operators must span at least the code distance $\delta$, incurring a free energy cost proportional to $\delta$. This leads to the AQEC relative entropy scaling as  $S(\Lambda+B\parallel\Lambda)\sim e^{-\delta /\xi}$, where $\xi$ denotes the correlation length related to the error rate. As the code distance increases with $n$, the relative entropy decreases.

In the disordered phase, some logical classes $l$ maintain a finite free energy cost, preventing $S(\Lambda + B || \Lambda)$ from reaching zero. For example, if $\Delta(\eta, l) \sim 0$ for all $l$, then $S(\Lambda + B || \Lambda) \sim 2k \log d$. Intuitively, when above the threshold, AQEC relative entropy characterizes how many logical qudits suffer from logical errors.

At a critical error threshold, the divergence of $\xi$ implies a power-law behavior $S(\Lambda + B || \Lambda) \sim 1/\delta^{2h}$. For instance, in the random bond Ising model, which describes the toric code and single qubit errors, the dual relationship between an open-ended wall (disorder operator) and spin-spin correlation exhibits a power-law length dependence. This can be extended to a closed domain wall $l$, leading to $e^{-\Delta} \sim 1/\delta^{2h}$, where $h$ is the Ising spin's scaling dimension.

For another example, assume we are dealing with a qudit LDPC code with $K=d^k$ and infinitely many logical qudits $k\rightarrow \infty$. Intuitively, AQEC relative entropy characterizes how many logical qudits suffer from logical errors. Suppose that $\Delta(\eta,l) \sim 0$ for all $l$, we have a divergent $S(\Lambda+B||\Lambda) \sim 2k \log d$ and $1-F_e \geq 1/2$, $1-\Pr(l=0) \geq 1/2$ from Eq. \eqref{eq:fidelity} and \eqref{eq:succ-prob}. But if $\Delta(\eta,l) \sim 0$ only for a finite number of logical classes $l$ and otherwise $\Delta(\eta,l) \rightarrow +\infty$, we get $S(\Lambda+B||\Lambda) \sim c \log d$ where $c$ is the number of failed logical classes.
So despite there being an incompatibility between AQEC relative entropy and fidelity or probability measures in Thm. \ref{thm1} and Thm. \ref{thm2} above the threshold, the AQEC relative entropy can still be a proper measure, while it is stronger in determining the recoverable region.

\section{Imperfect state preparation}
\label{sec:weak}
In the preceding section, we addressed the case using AQEC condition arising solely from noise channel $\mathcal{N}$. Yet, inevitable imperfections in the fundamental encoding channel $\mathcal{E}$ result in AQEC codes. We now illustrate how our formalism encompasses imperfect encoding or state preparation.

\begin{figure}
    \includegraphics[width=1\columnwidth]{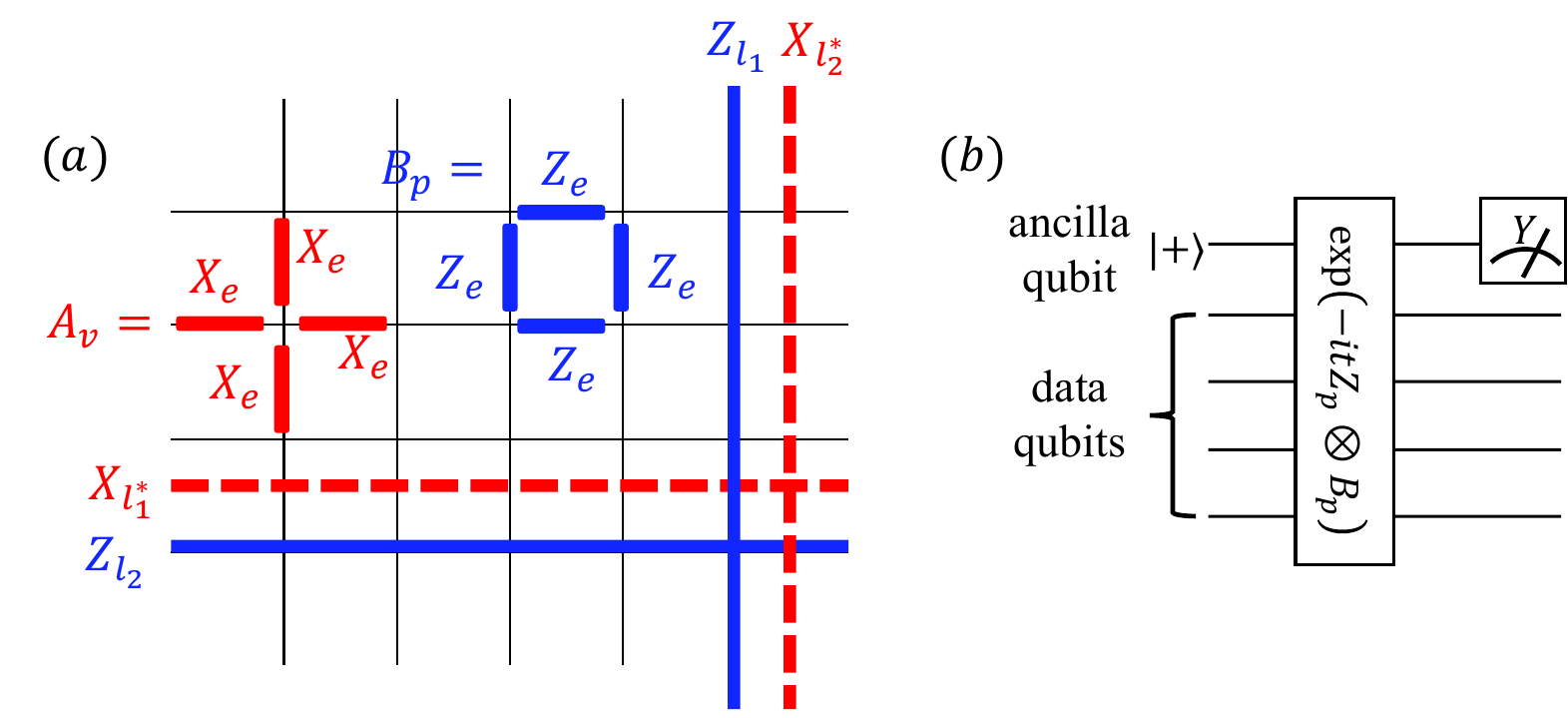}
    \caption{(a) Toric code defined on 2D periodic lattice. Physical qubits stay on the edges of the lattice. The two kinds of stabilizers are $A_{v}$ defined on each vertex and $B_{p} $ defined on each plaquette as shown in the figure. The logical Pauli $Z$ operators $Z_{l_1}$ and $Z_{l_2}$ are product of $Z$'s along non-contractible loops. Correspondingly the logical Pauli $X$ operators $X_{l_1^*}$ and $X_{l_2^*}$ are defined as $X$'s along non-contractible loops on the dual lattice.
    (b) A circuit model of $B_{p_0}$ measurement circuit.}
    \label{fig2}
\end{figure}

We consider toric code \cite{ref:dennis,KITAEV20032} on a 2D periodic lattice. The data qubits are located at edges $e$, and the stabilizer generators are
\begin{equation}
    A_{v}=\prod_{e|v\in\partial e}X_{e},\quad B_{p} = \prod_{e\in \partial p} Z_{e},
\end{equation}
as in Fig. \ref{fig2} (a). Normally the initial code states are prepared by measuring the stabilizer generators on a product state and projecting it onto the $+1$ subspace, for example,
\begin{equation}
    \ket{++}_L = \prod_{p} \frac{I+B_p}{2} \ket{+}^{\otimes n}.
\end{equation}
However, the measurement procedure might suffer from imperfection in reality. We need to entangle the data qubits with an ancilla qubit and then measure the ancilla in order to perform a four-qubit measurement, but the entanglement gate might acquire coherent error, leading to general positive operator-valued measurement (POVM) rather than projective measurement.
We consider the following model for imperfect measurement \cite{ref:guoyi,ref:fisher,zhao2023lattice} as in Fig. \ref{fig2} (b),
\begin{enumerate}
    \item prepare the ancilla qubit in $\ket{+}$ state for each plaquette $p$;
    \item apply a joint time evolution involving each ancilla and its four neighboring data qubits $\exp [-it Z_{p} \otimes B_{p}]$ where $Z_{p}$ is the Pauli $Z$ acting on ancilla at ${p}$ and we assume $0\leq t \leq \pi/4$;
    \item perform projective measurement on ancilla in $Y$ basis.
\end{enumerate}
The resulting measurement operator on data qubits is no longer a projection operator but instead \cite{ref:guoyi, zhao2023lattice}
\begin{equation}
    M_{\{s_{p}\}} =  \frac{1}{(\sqrt{2\cosh{\beta}})^{\frac{n}{2}}}\exp\left[\frac{1}{2}\beta \sum_{p} s_{p} B_{p}\right],
    \label{eq:imperfect-measurement}
\end{equation}
up to an irrelevant phase factor. Such measurement is also called weak measurement.
Here $s_p=\pm$ denotes the binary outcomes for every plaquette $p$ and $\tanh\beta/2 = \tan t$. The projective measurement is recovered when $t \rightarrow \pi/4$ or $\beta \rightarrow +\infty$.
We therefore assume the imperfect code subspace is spanned by (fixing the position of logical operators)
 \begin{equation}
\begin{aligned}
    &\ket{++}_L =\frac{M_{\{s_p=+\}}\ket{+}^{\otimes n} }{\sqrt{\bra{+}^{\otimes n} M_{\{s_p=+\}}^\dagger M_{\{s_p=+\}} \ket{+}^{\otimes n}}} \\&\propto \exp\left[\frac{1}{2}\beta B_p\right] \ket{+}^{\otimes n},\\&
    \ket{-+}_L=Z_{l_1} \ket{++}_L, \quad \ket{+-}_L=Z_{l_2}\ket{++}_L,\\& \ket{--}_L=Z_{l_1}Z_{l_2}\ket{++}_L.
\end{aligned}
\label{eq:preparation}
\end{equation}
It is able to verify that the above four states are orthogonal to each other.
This model is a rather simplified one which is easier to study analytically, but it can capture the fundamental influence of imperfect stabilizer measurement on QEC. In Ref. \cite{ref:guoyi,ref:fisher} it is shown that these states lose long-range entanglement or topological order, and in Ref. \cite{zhao2023lattice} we showed that it is undecodable under single-qubit Pauli $X$ noises through a common multi-round syndrome measurement protocol. Here we analyze this model using AQEC condition and AQEC relative entropy to extract an optimal threshold.

We want to examine its ability of protecting logical information against single-qubit bit-flip noise channel,
\begin{equation}
    \mathcal{N}(\rho) = \prod_{e} \mathcal{N}_e (\rho), \quad \mathcal{N}_e (\rho) = (1-p) \rho + p X_e \rho X_e.
\end{equation}
Unlike the perfect preparation case, $B_{uv,q_1q_2}$ is now nonzero even for certain local noises like $E^\dagger_u E_v \sim \sqrt{p} X_e$ and scales as $\mathcal{O}\left(p^{\frac{1}{2}} e^{-2\beta}\right)$ for large enough $\beta$ and small enough $p$, which is not suppressed by code distance(see Appendix \ref{sec:SIII}). It suggest that as long as $\beta$ is finite, $B_{uv,q_1q_2}$ is always comparable to $\Lambda_{uv,q_1q_2}$ and prevents $S(\Lambda+B||\Lambda)$ from approaching $0$ in the thermodynamic limit.

We can still try to find an SM interpretation. In general, this is not always valid beyond stabilizer code states and Weyl errors due to the violation of Gottesman-Knill theorem, but luckily it works for the current simple model Eq.\eqref{eq:preparation}. We write $\tr\left( (\Lambda+B)^R\right)$ as a partition function
\begin{equation}
    \begin{aligned}
        \tr\left( (\Lambda+B)^R\right) &\propto \sum_{\{\eta^{(\alpha)}_e\}} \exp[-H(\{\eta^{(\alpha)}_e\})],\\
        H(\{\eta^{(\alpha)}_e\}) &= -\sum_{\alpha=1}^R \left[h  \sum_{e}\eta^{(\alpha)}_e \right.\\
        &+ \frac{1}{2} \log \cosh \beta  \sum_{p}U^{(\alpha)}_p U^{(\alpha+1)}_p \\
        &\left.+ \log \left(1 + (\tanh \beta)^{n/2} \prod_{p} \delta_{U^{(\alpha)}_p, U^{(\alpha+1)}_p}\right)\right],
    \end{aligned}
    \label{eq:partition2}
\end{equation}
Where $h=\frac{1}{2}\log\frac{1-p}{p}$ and $U^{(\alpha)}_p = \prod_{e\in \partial p }\eta^{(\alpha)}_e$. The SM d.o.f. $\eta^{(\alpha)}_e = \pm 1$ is defined on each data qubit $e$ and each replica copy $\alpha=1,\cdots,R$ (we identify $\eta^{(R+1)}_e=\eta^{(1)}_e$). $\{\eta^{(\alpha)}_e\}$ correspond to the error configurations $u$, $v$ in Eq. \eqref{eq:gKL}, thus the above partition function captures fluctuations of errors.
The phase transition point for $R\rightarrow 1$ limit is identified as the intrinsic threshold.


\begin{figure}
    \includegraphics[width=1\columnwidth]{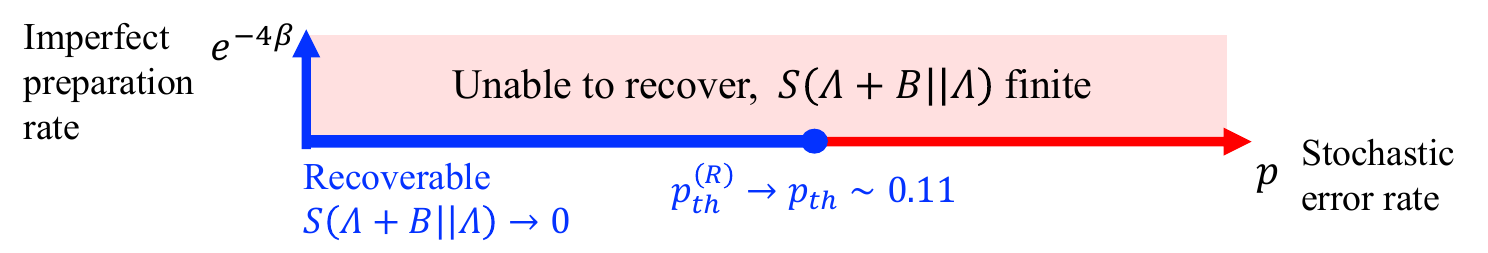}
    \caption{The phase diagram. The undecodable phase stays at any finite preparation and Pauli error rates, as well as the region $e^{-4\beta}=0$ and $p > p_{th}$. Notice that the RBIM phase transition point will be different for different replica $R$, and $p_{th}$ is obtained in $R\rightarrow 1$.}
    \label{fig3}
\end{figure}
The R\'{e}nyi AQEC relative entropy can be expressed as an observable of the SM model, 
\begin{equation}
\begin{aligned}
        S^{(R)}(\Lambda+B||\Lambda) &= \frac{1}{R-1} \log \frac{\tr((\Lambda+B)^R)}{\tr(\Lambda^{R})}\\& = \frac{1}{1-R} \log \Braket{\prod_{\alpha,l} \frac{1+\prod_{e\in l} \eta^{(\alpha)}_e\eta^{(\alpha+1)}_e}{2}},
\end{aligned}
\end{equation}
allowing further calculations. As a result,
when $\beta \rightarrow +\infty$ the model Eq. \eqref{eq:partition2} reduces to Eq. \eqref{eq:partition1}, or more concretely replica random bond Ising model \cite{ref:dennis,nishimoriStatisticalPhysicsSpin2001} with domain wall inserted.
When $\beta \rightarrow 0$ and keeping $h$ finite, we obtain a trivial paramagnetic spin model and $\lim_{n\rightarrow \infty}S^{(R)}(\Lambda+B||\Lambda) = 2\log 2$. We then focus on the case when both state preparation and Pauli error rates are small but nonzero $0 < e^{-\beta},p \ll 1$ and perturbatively calculate $S^{(R)}(\Lambda+B||\Lambda)$ up to the first non-vanishing order (see Appendix \ref{sec:SIII}), 
\begin{equation}
    S^{(R)}(\Lambda+B||\Lambda) \sim \frac{\sqrt{n/2}}{1-1/R}  \exp\left(-2h-4\beta\right).
    \label{eq:expansion}
\end{equation}
The coefficient $\sqrt{n}$ suggests that the imperfect code cannot suppress $S^{(R)}(\Lambda+B||\Lambda)$ to $0$ by increasing $n$. Consequently the phase transition is located at $1/\beta=0$ for any $R>1$ (detailed discussions are in Appendix \ref{sec:SIII}).
Then we extrapolate to $R\rightarrow 1$ and conclude that the intrinsic threshold is at $e^{-4\beta}=0$ or $\beta \rightarrow +\infty$. In other words, the parameter region below the intrinsic threshold is $1/\beta=0$ and $p< p_c \sim 0.11$, as shown in the phase diagram Fig. \ref{fig3}. Our theoretical framework correctly conveys the insight that information encoded in  imperfectly prepared toric code \eqref{eq:preparation} subjected to single-qubit Pauli errors is irrecoverable, regardless of the decoding strategy employed. This is also compatible with our previous result in Ref. \cite{zhao2023lattice}, where we assumed a specific noisy decoding procedure.


\section{Discussion}
Note that the last example confirmed that our framework could be applied on general AQEC codes, which possess nonzero $B_{uv,q_1q_2}$ for local error operators $E_u^\dagger E_v$. Although the intrinsic threshold in this example vanishes due to the non-vanishing $B_{uv,q_1q_2}$ in the $n\rightarrow \infty$ limit, it may remain finite for certain AQEC codes with $B_{uv,q_1q_2}$ suppressed by size (potentially e.g. topological ordered states \cite{bravyiTopologicalQuantumOrder2010}, approximate quantum LDPC code \cite{BohdanowiczLDPC}, ETH and Heisenberg chain codes \cite{BrandaoETH}). 
Our framework holds potential for application to codes and noises that are beyond Gottesman-Knill theorem, including above AQEC codes and amplitude damping noises \cite{darmawan_tensor}.




\begin{acknowledgments}
The authors thank Zi-wen Liu and Ying-fei Gu for valuable discussions. The work is supported by the National Natural Science Foundation of China (Grant No.~92365111), Beijing Natural Science Foundation (Grants No.~Z220002), and the Innovation Program for Quantum Science and Technology (Grant No.~2021ZD0302400).
\end{acknowledgments}

\begin{appendix}


\setcounter{lemma}{0}
\section{entanglement fidelity and coherent information}
\label{sec:SI}

\begin{figure}[h]
    \includegraphics[width=0.5\columnwidth]{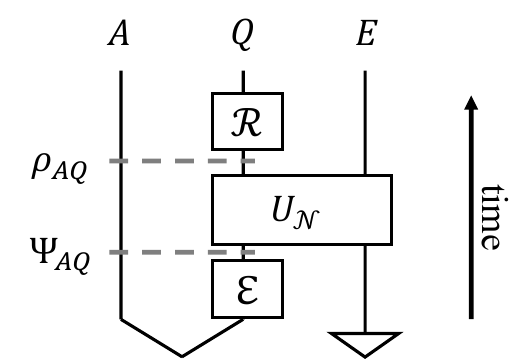}
    \caption{The circuit that underlies the definition of coherent information and entanglement entropy. Here $\mathcal{E}$ is the encoding channel, $\mathcal{R}$ is the recovery channel, $U_{\mathcal{N}}$ comes from purifying noise channel.}
    \label{fig1}
\end{figure}
Suppose that we have a state $\rho_{0}$ storing logical information and we encode it in the code subspace $\mathcal{C}$. $Q$ labels the physical system of the code. We introduce a reference system $A$ and purify it as $\ket{\Psi_{AQ}} = \sum_{q_1,q_2} (\sqrt{\rho_0})_{q_1 q_2} \ket{q_1}_A \ket{q_2}_{\mathcal{C}}$, where $\ket{q}_\mathcal{C}$'s are the logical code words in the logical space and $\ket{q}_A$'s are the corresponding ancilla states. The dimension of Hilbert space of $A$ is the code subspace dimension $K$. The system $Q$ suffers from noise channel $\mathcal{N}(\rho) = \sum_{u} E_u \rho E_u^\dagger$, we also purify it by introducing an environment $E$ as in Fig. \ref{fig1}. The post-error pure state reads
\begin{equation}
    \ket{\Psi_{AQE}'} = \sum_{u} E_u \ket{\Psi_{AQ}} \otimes \ket{u}_E,
\end{equation}
such that $\rho_{AQ} =\sum_{u} E_u \ket{\Psi_{AQ}}\bra{\Psi_{AQ}} E_u^\dagger$ after tracing out $E$.  The coherent information is defined as 
\begin{equation}
    I_c(\rho_{0},\mathcal{ N \circ E}) = S(\rho_Q)-S(\rho_{AQ}),
    \label{eq:coherent-app}
\end{equation}
where $S(\rho) = -\tr\rho\log\rho$ is the von Neumann entropy and $\rho_Q = \tr_A \rho_{AQ}$. It satisfies
\begin{equation}
    -S(\rho_A)\leq I_c(\rho_{0},\mathcal{ N \circ E})\leq S(\rho_A),
    \label{eq:range}
\end{equation}
and characterizes how much entanglement between $A$ and $Q$ is preserved under the noise channel. Assume that we apply a recovery $\mathcal{R}$ on $Q$, The entanglement fidelity is defined as
\begin{equation}
    F_e (\rho_{0},\mathcal{R\circ N \circ E}) = \bra{\Psi_{AQ}}\mathcal{R}(\rho_{AQ}) \ket{\Psi_{AQ}},
    \label{eq:ef}
\end{equation}
which quantifies how much quantum information is protected after the entire QEC process. 
The Nielsen-Schumacher condition for exact recovery $\mathcal{R\circ N \circ E} = \mathrm{id}$ is $I_c(\rho_{0},\mathcal{ N \circ E}) = S(\rho_A)$, which is equivalent to the Knill-Laflamme (KL) condition. Actually, it is sufficient to choose $\rho_{0}$ as the maximumly mixed state $I_K/K$ to probe exact recovery. In the following sections, we set $\rho_{0}=I_K/K$ such that the quantities \eqref{eq:coherent-app} and \eqref{eq:ef} can be viewed as 'averaged' over all logical states, instead of taking a maximization or minimization as in Ref. \cite{tyson_two-sided_2010,beny_general_2010}.
Now the state $\ket{\Psi_{AQ}}$ can be expressed as
\begin{equation}
    \ket{\Psi_{AQ}} = \frac{1}{\sqrt{K}}\sum_{q} \ket{q}_A \otimes \ket{q}_\mathcal{C}.
\end{equation}
Similarly,
\begin{equation}
    \rho_{0Q} \equiv \tr_{A} \ket{\Psi_{AQ}}\bra{\Psi_{AQ}} = \frac{1}{K}  \ket{q}_\mathcal{C}\bra{q}_\mathcal{C} = \frac{1}{K} P,
\end{equation}
where $P$ is the projection on to the code subspace $\mathcal{C}$.
We abbreviate that $I_c = I_c(I_K/K,\mathcal{ N \circ E})$ and $F_e (\mathcal{R\circ N \circ E}) = F_e (I_K/K,\mathcal{R\circ N \circ E})$.

\begin{lemma}
\label{lemma1-app}
For a QEC code $\mathcal{C}$ under a noise channel $\mathcal{N}$, the AQEC relative entropy is related to the coherent information through:
\begin{equation}
    S(\Lambda+B||\Lambda) = -I_c+ \log K.
\end{equation}
\end{lemma}
\begin{proof}

Since $\ket{\Psi_{AQE}'}$ is a pure state, the coherent information has an alternative form
\begin{equation}
    I_c = S(\rho_{AE}) - S(\rho_E),
\end{equation}
where 
\begin{align}
         \rho_{E} &= \tr_{AQ} \ket{\Psi_{AQE}'} \bra{\Psi_{AQE}'} \notag\\
        &= \sum_{uv} \tr_{AQ}\left( E_u \ket{\Psi_{AQ}}\bra{\Psi_{AQ}} E_v^\dagger \right) \ket{u}_E \bra{v}_E \notag\\
        &= \frac{1}{K}\sum_{uv} \tr_{Q}\left( E_u P E_v^\dagger \right) \ket{u}_E \bra{v}_E \notag\\
        &= \frac{1}{K}\sum_{uv} \tr_{Q}\left(P E_v^\dagger E_u P \right) \ket{u}_E \bra{v}_E \notag\\
        &= \frac{1}{K}\sum_{uv} \tr_{Q}\left(\lambda_{vu} P + PB_{vu} P \right) \ket{u}_E \bra{v}_E \notag\\
        &= \sum_{uv} \lambda_{vu}  \ket{u}_E \bra{v}_E.
\end{align}
Notice that we have used $\tr_{\mathcal{C}}(B_{vu})=0$. Similarly,
\begin{align}
     \rho_{AE} &= \tr_{Q} \ket{\Psi_{AQE}'} \bra{\Psi_{AQE}'} \notag\\
        &= \sum_{uv}\left[\tr_{Q}\left( E_u \ket{\Psi_{AQ}}\bra{\Psi_{AQ}} E_v^\dagger \right)\right] \ket{u}_E \bra{v}_E \notag\\
        &= \frac{1}{K}\sum_{uv,q_1q_2}\left[ \tr_{Q}\left( E_u \ket{q_1}_\mathcal{C}\bra{q_2}_\mathcal{C} E_v^\dagger \right) \right] \notag\\
        &\times \ket{u}_E \bra{v}_E \otimes \ket{q_1}_A\bra{q_2}_A \notag\\
        &= \frac{1}{K}\sum_{uv,q_1q_2} \bra{q_2}_\mathcal{C} E_v^\dagger E_u \ket{q_1}_\mathcal{C} \ket{u}_E \bra{v}_E \otimes \ket{q_1}_A\bra{q_2}_A \notag\\
        &= \sum_{uv} (\Lambda_{vu,q_2q_1}+B_{vu,q_2q_1})  \ket{u}_E \bra{v}_E \otimes \ket{q_1}_A\bra{q_2}_A.
\end{align}
Since $\rho_{A}=I_K/K$, we obtain the relation between the generalized KL condition and the error correction circuit Fig. \ref{fig1},
\begin{equation}
    \Lambda = (\rho_E\otimes \rho_A)^T,\quad \Lambda+B= \rho_{AE}^T.
\end{equation}
The density matrix nature of the $\Lambda$ and $\Lambda+B$ matrices tells us that they are both positive semi-definite $\Lambda \geq 0$,  $\Lambda+B\geq 0$ and trace-one $\tr(\Lambda)=1$, $\tr(\Lambda+B)=1$.
Thus the AQEC relative entropy is expressed as
\begin{equation}
    \begin{aligned}
        S(\Lambda+B||\Lambda) &= \tr\left\{(\Lambda+B)\left[\log(\Lambda+B)-\log\Lambda\right]\right\}\\
        &=\tr(\Lambda+B)\log(\Lambda+B)-\tr\Lambda\log\Lambda\\
        &=-S(\rho_{AE})+S(\rho_A)+S(\rho_E)\\
        &=-I_c+\log K,
    \end{aligned}
\end{equation}
where we used $\tr (B\log(\Lambda))=0$ which is derived from $\tr_{\mathcal{C}}(B)=0$. More specifically, note that the $\Lambda$ matrix is diagonal in the code subspace, 
\begin{equation}
    \Lambda_{uv,q_1q_2}=\lambda_{uv} \delta_{q_1q_2}/K,
\end{equation}
where $q_1,q_2$ are code words and $u,v$ are error configurations. In particular, $\Lambda$ has a tensor product structure,
\begin{equation}
    \Lambda = \lambda \otimes I_K/K.
\end{equation}
$\lambda$ acts only on the error configuration space and $I_K$ is the identity in the code subspace $\mathcal{C}$.
The logarithm becomes $\log(\Lambda) = \log(\lambda/K)\otimes I_K$. So,
\begin{equation}
\begin{aligned}
        \tr_{\mathcal{C}} (B\log(\Lambda)) &= \tr_{\mathcal{C}} (B\log(\lambda/K)\otimes I_K)\\& = \tr_{\mathcal{C}} (B)\log(\lambda/K) =0,
\end{aligned}
\end{equation}
which means $\tr (B\log(\Lambda))=0$.
    
\end{proof}
Note that the AQEC relative entropy takes value from
\begin{equation}
    0\leq S(\Lambda+B||\Lambda)\leq 2\log K,
    \label{eq:range2}
\end{equation}
following Eq. \eqref{eq:range}.

\section{Proof of Lemma \ref{lemma2}}\label{sec:SII}

\begin{lemma}
\label{lemma2-app}
    For a qudit stabilizer code in prime $d$ and a stochastic Weyl noise channel, the AQEC relative entropy, probability of logical classes and order parameter of the SM model are related by
    \begin{equation}
    \begin{aligned}
        S(\Lambda+B||\Lambda) &=H(L|S)\\&= \sum_{\eta \in \mathcal{V}} \Pr(\eta) \log \left\{ \sum_{l \in \mathcal{L}} \exp\left[ -\Delta(\eta,l)\right]\right\},
    \end{aligned}
    \label{eq:main-app}
    \end{equation}
    where $H(L|S)$ is the Shannon conditional entropy of logical class $l$ given syndrome $s$, and the last quantity is a generalized version of the homological difference  \cite{kovalevNumericalAnalyticalBounds2018}.
\end{lemma}

\begin{proof}
First, we evaluate the expression for the matrices $\Lambda$ and $\Lambda+B$.
Consider the stochastic Weyl error
\begin{equation}
    \mathcal{N} (\rho) = \sum_{u \in \mathcal{V}} \Pr(u) T(u) \rho T(u)^\dagger,
\end{equation}
where $\Pr(u)$ is a probabilistic distribution on $\mathcal{V}$, $0\leq \Pr(u) \leq 1$, $ \sum_{u \in \mathcal{V}} \Pr(u) = 1$.
Now $P$ is the projection onto the stabilizer code subspace $\mathcal{C}$ and the Kraus operator of stochastic Weyl error channel reads $E_u = \sqrt{\Pr(u)} T(u)$. Then we have
\begin{align}
        &P E^\dagger_u E_v P\notag\\
        & = \frac{1}{d^{2r}} \sum_{m,m'\in \mathcal{M}} \sqrt{\Pr(u)\Pr(v)} T(m)T(u)^\dagger T(v) T(m') \notag\\
        &=  \frac{1}{d^{2r}} \sum_{m,m'\in \mathcal{M}} \sqrt{\Pr(u)\Pr(v)}\omega^{-\frac{1}{2}[u,v]} \omega^{[m,v-u]}\notag\\
        &\times T(v-u) T(m+m')\notag\\
        &= \frac{1}{d^{r}} \sum_{m''\in \mathcal{M}} \sqrt{\Pr(u)\Pr(v)}\omega^{-\frac{1}{2}[u,v]}\notag\\
        &\times \delta_{\mathcal{M}^\perp}(v-u) T(v-u) T(m'')\notag\\
        &= \sqrt{\Pr(u)\Pr(v)}\omega^{-\frac{1}{2}[u,v]} \delta_{\mathcal{M}^\perp}(v-u) T(v-u) P\notag\\
        &= \sqrt{\Pr(u)\Pr(v)}\omega^{-\frac{1}{2}[u,v]} \delta_{\mathcal{M}}(v-u) P\notag\\
        &+ \sqrt{\Pr(u)\Pr(v)}\omega^{-\frac{1}{2}[u,v]} \delta_{\mathcal{M}^\perp - \mathcal{M}}(v-u) P T(v-u) P,
\end{align}

where we have used Lemma 9 in Ref. \cite{gross}
\begin{equation}
    \frac{1}{|\mathcal{M}|} \sum_{m\in \mathcal{M}} \omega^{[v,m]} = \delta_{\mathcal{M}^\perp}(v) = \left\{ \begin{matrix} 1 &v\in \mathcal{M}^\perp\\0 & \text{else.}\end{matrix} \right. 
\end{equation}
Here $\delta_{\mathcal{M}^\perp - \mathcal{M}}(v-u)=1$ only when $v-u \in \mathcal{M}^\perp$ but $\notin \mathcal{M}$. When $v-u\in \mathcal{M}$ the Weyl operator $T(v-u)$ is just a stabilizer and absorbed into $P$, which leads to the $\lambda_{uv}$ term. When $v-u \in \mathcal{M}^\perp$ but $\notin \mathcal{M}$, $T(v-u)$ is a logical operator which causes logical error, hence it corresponds to the $B_{uv}$ term.
Therefore,
\begin{equation}
    \begin{aligned}
        &\lambda_{uv} = \sqrt{\Pr(u)\Pr(v)}\omega^{-\frac{1}{2}[u,v]} \delta_{\mathcal{M}}(v-u),\\
        &B_{uv} = \sqrt{\Pr(u)\Pr(v)}\omega^{-\frac{1}{2}[u,v]} \delta_{\mathcal{M}^\perp - \mathcal{M}}(v-u) T(v-u).
    \end{aligned}
\end{equation}
We will show that $B_{uv}$ is suppressed in the ordered phase in an average sense.

In order to calculate the AQEC relative entropy, we apply the replica method. Owing to $\tr_{\mathcal{C}}(B)=0$, we have 
\begin{equation}
\begin{aligned}
        S(\Lambda+B||\Lambda) &= \tr(\Lambda+B)\log(\Lambda+B) - \tr\Lambda\log \Lambda \\&= S(\Lambda) - S(\Lambda+B),
\end{aligned}
\end{equation}
which is the difference between the two von Neumann entropies $S(\Lambda)$ and $S(\Lambda+B)$. We extend both of them to R\'{e}nyi entropies, 
\begin{equation}
\begin{aligned}
    &S^{(R)}(\Lambda+B||\Lambda) = S^{(R)}(\Lambda) - S^{(R)}(\Lambda+B) \\
     &S^{(R)}(\rho) = \frac{1}{1-R} \log \tr(\rho^{R}),
\end{aligned}
\end{equation}
which is holomorphic in $R\in \mathbb{C}$ for $\Re R \geq 1$ \cite{wittenOpenStringsRindler2019}. The von Neumann entropy can be extracted from the R\'{e}nyi entropy through the $R \rightarrow 1$ limit,
\begin{equation}
    S(\rho) = \lim_{R\rightarrow 1} S^{(R)}(\rho) = S^{(1)}(\rho) = -\left. \frac{\d}{\d R} \tr(\rho^{R}) \right|_{R\rightarrow 1}.
\end{equation}

Now we compute the R\'{e}nyi entropies for integer $R>1$,
then carefully perform analytic continuation and take the $R\rightarrow 1$ limit $\lim_{R\rightarrow 1} S^{(R)}(\Lambda+B||\Lambda)=S(\Lambda+B||\Lambda)$. We first compute $\tr(\Lambda^R)$,
\begin{widetext}
\begin{align}
        \tr(\Lambda^R) &= \frac{1}{d^{k(R-1)}} \left(\prod_{\alpha=1}^R \sum_{\eta^{(\alpha)} \in \mathcal{V}} \right) \prod_{\alpha=1}^R \lambda_{\eta^{(\alpha)}\eta^{(\alpha+1)}}\notag\\
        &=\frac{1}{d^{k(R-1)}} \left(\prod_{\alpha=1}^R \sum_{\eta^{(\alpha)} \in \mathcal{V}} \right) \prod_{\alpha=1}^R \sqrt{\Pr(\eta^{(\alpha)})\Pr(\eta^{(\alpha+1)})}\omega^{-\frac{1}{2}[\eta^{(\alpha)},\eta^{(\alpha+1)}]} \delta_{\mathcal{M}}(\eta^{(\alpha+1)}-\eta^{(\alpha)})\notag\\
        &=\frac{1}{d^{k(R-1)}} \left(\sum_{\eta\in \mathcal{V}} \prod_{\alpha=1}^{R-1} \sum_{m^{(\alpha)} \in \mathcal{M}} \right)  \Pr(\eta) \prod_{\alpha=1}^{R-1} \Pr(\eta+m^{(\alpha)}) \prod_{\alpha=1}^{R} \omega^{-\frac{1}{2}[\eta+m^{(\alpha)},\eta+m^{(\alpha+1)}]}\notag\\
        &=\frac{1}{d^{k(R-1)}} \left(\sum_{\eta\in \mathcal{V}} \prod_{\alpha=1}^{R-1} \sum_{m^{(\alpha)} \in \mathcal{M}} \right)  \Pr(\eta) \prod_{\alpha=1}^{R-1} \Pr(\eta+m^{(\alpha)}) \prod_{\alpha=1}^{R} \omega^{-\frac{1}{2}[\eta,m^{(\alpha+1)}] + \frac{1}{2}[\eta,m^{(\alpha)}]}\notag\\
        &=\frac{1}{d^{k(R-1)}} \left(\sum_{\eta\in \mathcal{V}} \prod_{\alpha=1}^{R-1} \sum_{m^{(\alpha)} \in \mathcal{M}} \right)  \Pr(\eta) \prod_{\alpha=1}^{R-1} \Pr(\eta+m^{(\alpha)})\notag\\
        &=\frac{1}{d^{k(R-1)}} \sum_{\eta\in \mathcal{V}}  \Pr(\eta)\left[ \sum_{m \in \mathcal{M}}  \Pr(\eta+m)\right]^{R-1}.
    \label{eq:trL}
\end{align}
Here $\alpha$ denotes the replica index that comes from the $R$ copies of $\Lambda$ matrix, and we identify $\eta^{(R+1)}=\eta^{(1)}$. $\eta^{(\alpha)}$ is the error configuration for the copy $\alpha$.
In the third line, notice that $\prod_{\alpha =1}^R \delta_{\mathcal{M}}(\eta^{(\alpha + 1)} - \eta^{(\alpha)}) = \prod_{\alpha =1}^{R-1} \delta_{\mathcal{M}}(\eta^{(\alpha)} - \eta^{(R)}) $ demands that all the $\eta^{(\alpha)}$s differ from $\eta^{(R)}$ by element in $\mathcal{M}$, i.e. a stabilizer, so we denoted $\eta^{(R)} = \eta$ and substituted $m^{(\alpha)} = \eta^{(\alpha)} - \eta$ for $\alpha = 1, \cdots, R-1$. The $m^{(\alpha)}$ variables are restricted in $\mathcal{M}$ by the $\delta$ symbol. So the summation $\prod_{\alpha=1}^R \sum_{\eta^{(\alpha)} \in \mathcal{V}}$ reduce to $\sum_{\eta\in \mathcal{V}} \prod_{\alpha=1}^{R-1} \sum_{m^{(\alpha)} \in \mathcal{M}}$. In the fourth line, we notice that the product of phase factors yields $1$, leading to the fifth line.

Similarly we compute $\tr((\Lambda+B)^R)$,
\begin{equation}
    \begin{aligned}
        \tr((\Lambda+B)^R)
        &=\frac{1}{d^{kR}} \left(\prod_{\alpha=1}^R \sum_{\eta^{(\alpha)} \in \mathcal{V}} \right) \prod_{\alpha=1}^R \sqrt{\Pr(\eta^{(\alpha)})\Pr(\eta^{(\alpha+1)})}\omega^{-\frac{1}{2}[\eta^{(\alpha)},\eta^{(\alpha+1)}]} \delta_{\mathcal{M}^\perp}(\eta^{(\alpha+1)}-\eta^{(\alpha)}) \tr_\mathcal{C}\left(\prod_{\alpha=1}^R T(\eta^{(\alpha+1)}-\eta^{(\alpha)})\right)\\
        &=\frac{1}{d^{kR}} \left(\sum_{\eta\in \mathcal{V}} \prod_{\alpha=1}^{R-1} \sum_{m^{(\alpha)} \in \mathcal{M}} \sum_{l^{(\alpha)} \in \mathcal{L}}\right)  \Pr(\eta) \prod_{\alpha=1}^{R-1} \Pr(\eta+m^{(\alpha)}+l^{(\alpha)}) \prod_{\alpha=1}^{R} \omega^{-\frac{1}{2}[\eta+m^{(\alpha)}+l^{(\alpha)},\eta+m^{(\alpha+1)}+l^{(\alpha+1)}]}\\
        &\times \tr_\mathcal{C}\left(\prod_{\alpha=1}^R T(m^{(\alpha+1)}+l^{(\alpha+1)}-m^{(\alpha)}-l^{(\alpha)})\right)\\
        &=\frac{1}{d^{kR}} \left(\sum_{\eta\in \mathcal{V}} \prod_{\alpha=1}^{R-1} \sum_{m^{(\alpha)} \in \mathcal{M}} \sum_{l^{(\alpha)} \in \mathcal{L}}\right)  \Pr(\eta) \prod_{\alpha=1}^{R-1} \Pr(\eta+m^{(\alpha)}+l^{(\alpha)}) \prod_{\alpha=1}^{R} \omega^{-\frac{1}{2}[l^{(\alpha)},l^{(\alpha+1)}]}\\
        &\times \tr_\mathcal{C}\left(\prod_{\alpha=1}^R T(m^{(\alpha+1)}+l^{(\alpha+1)}-m^{(\alpha)}-l^{(\alpha)})\right).\\
    \end{aligned}
\end{equation}
\end{widetext}
Here $\tr_\mathcal{C}(*) = \sum_{q}\bra{q}_{\mathcal{C}} * \ket{q}_{\mathcal{C}}$ denotes the trace over the code subspace.
In the second line we eliminated the $\delta$ symbol and performed variable substitution as in Eq. \eqref{eq:trL}. Notice that here $\prod_{\alpha =1}^R \delta_{\mathcal{M}^\perp}(\eta^{(\alpha + 1)} - \eta^{(\alpha)}) = \prod_{\alpha =1}^{R-1} \delta_{\mathcal{M}^\perp}(\eta^{(\alpha)} - \eta^{(R)})$ now requires the $\eta^{(\alpha)}$ differs from each other by a element in $\mathcal{M}^\perp$. Recall Eq. \eqref{eq:Mperp}, $\mathcal{M}^\perp$ contains all possible choices of logical operators. Since $\mathcal{M}^\perp$ has the tensor product structure $\mathcal{M}^\perp= \mathcal{L} \otimes \mathcal{M}$, where $\mathcal{L}$ is the space of logical classes and $\mathcal{M}$ is the stabilizers, we chose the substitute the variables as $\eta^{(\alpha)} = m^{(\alpha)} + l^{(\alpha)} + \eta$, $m^{(\alpha)} \in \mathcal{M}$, $l^{(\alpha)} \in \mathcal{L}$ for $\alpha = 1, \cdots, R-1$ and $\eta^{(R)}=\eta$. The summation of the variables changes accordingly. In the third line, the exponents of  the phase factor involving $\eta$ and $m^{(\alpha)}$ cancel as in Eq. \eqref{eq:trL}, but the $\omega^{-\frac{1}{2}[l^{(\alpha)},l^{(\alpha+1)}]}$ term remains.
Now we evaluate the $\tr_\mathcal{C}$ term.
Using the fact stabilizers are symplectic orthogonal to logical operators $[m,l]=0$ and acts trivially on the logical states $T(m)\ket{q}_{\mathcal{C}}=\ket{q}_{\mathcal{C}}$, we obtain
\begin{equation}
    \begin{aligned}
        &\tr_\mathcal{C}\left(\prod_{\alpha=1}^R T(m^{(\alpha+1)}+l^{(\alpha+1)}-m^{(\alpha)}-l^{(\alpha)})\right)\\& = \tr_\mathcal{C}\left(\prod_{\alpha=1}^R T(l^{(\alpha+1)}-l^{(\alpha)})\right).\\
    \end{aligned}
\end{equation}
Then use the algebraic relation Eq. \eqref{eq:algebra}, we have
\begin{equation}
\begin{aligned}
        &\tr_\mathcal{C}\left(\prod_{\alpha=1}^R T(l^{(\alpha+1)}-l^{(\alpha)})\right)\\& = 
    \omega^{\frac{1}{2}\sum_{\alpha=2}^{R}[l^{(\alpha)}-l^{(1)},l^{(\alpha+1)}-l^{(\alpha)}]}
    \tr_\mathcal{C}\left( T\left(\sum_{\alpha=1}^R (l^{(\alpha+1)}-l^{(\alpha)})\right)\right)\\
    & = d^k
    \omega^{\frac{1}{2}\sum_{\alpha=1}^{R}[l^{(\alpha)},l^{(\alpha+1)}]}.\\
\end{aligned}
\end{equation}
Therefore we obtain
\begin{equation}
\begin{aligned}
        &\tr((\Lambda+B)^R) \\& =\frac{1}{d^{k(R-1)}} \left(\sum_{\eta\in \mathcal{V}} \prod_{\alpha=1}^{R-1} \sum_{m^{(\alpha)} \in \mathcal{M}} \sum_{l^{(\alpha)} \in \mathcal{L}}\right)  \Pr(\eta) \\&\times \prod_{\alpha=1}^{R-1} \Pr(\eta+m^{(\alpha)}+l^{(\alpha)})\\
        &=\frac{1}{d^{k(R-1)}} \sum_{\eta\in \mathcal{V}}  \Pr(\eta)\left[ \sum_{m \in \mathcal{M}} \sum_{l \in \mathcal{L}} \Pr(\eta+m+l)\right]^{R-1}.\\
\end{aligned}
\label{eq:trLB}
\end{equation}

The Equations \eqref{eq:trL} and \eqref{eq:trLB} now only hold for integer $R\geq 1$ (the case $R=1$ is trivial). We extend these equalities to $\Re R \geq 1$ by applying Carlson's theorem \cite{boas_entire_1973,wittenOpenStringsRindler2019,dhokerAlternativeMethodExtracting2021}. Specifically, for Eq. \eqref{eq:trL}, the last expression can be analytically continued to a holomorphic function on $\Re R \geq 1$,
\begin{equation}
    g(R)= \frac{1}{d^{k(R-1)}} \sum_{\eta\in \mathcal{V}}  \Pr(\eta)\left[ \sum_{m \in \mathcal{M}}  \Pr(\eta+m)\right]^{R-1}.
\end{equation}
For $\Re R \geq 1$, since $\Pr(\eta)$ is a probability distribution, we have $0 \leq \sum_{m \in \mathcal{M}}  \Pr(\eta+m) \leq 1 $ and thus 
\begin{equation}
    \left|\left[ \sum_{m \in \mathcal{M}}  \Pr(\eta+m)\right]^{R-1}\right| \leq 1.
\end{equation}
It follows that 
\begin{equation}
\begin{aligned}
     &\left|\sum_{\eta\in \mathcal{V}}  \Pr(\eta)\left[ \sum_{m \in \mathcal{M}}  \Pr(\eta+m)\right]^{R-1} \right| \\&\leq \sum_{\eta\in \mathcal{V}}  \Pr(\eta) \left|\left[ \sum_{m \in \mathcal{M}}  \Pr(\eta+m)\right]^{R-1} \right| \leq 1.        
\end{aligned}
\end{equation}
So the norm of $g(R)$ is bounded,
\begin{equation}
   | g(R)| \leq |d^{-k(R-1)}| \leq 1,
\end{equation}
for constants $d\geq 2$ and $k \geq 0$. 
For the l.h.s., we already know that $\tr(\Lambda^R)$ is bounded in absolute value by $1$ for $\Re R \geq 1$, $|\tr(\Lambda^R)| \leq 1$.
So $|\tr(\Lambda^R)-g(R)|$ is bounded by $2$. The difference $\tr(\Lambda^R)-g(R)$ satisfies the requirements of Carlson's theorem
\footnote{The Carlson's theorem states that given a holomorphic function $g(R)$ for $\Re R \geq 1$, it is identically zero, $g(R)=0, \forall \Re R \geq 1$, if the following requirements are satisfied: 1. $|g(R)|< C e^{\lambda_1 |R|}$ for some constant $C$, $\lambda_1$; 2. on the line $\Re R = 1$, $|g(R)|< C e^{\lambda_2 |R|}$ for some constant $\lambda_2 < \pi$; 3. $g(R)=0$ for positive integers $R$.}, thus we conclude that $\tr(\Lambda^R)=g(R)$ for $\Re R \geq 1$. A similar argument can be applied to Eq. \eqref{eq:trLB} such that it also holds for $\Re R \geq 1$.

The AQEC relative entropy is then obtained by taking the derivative at $R=1$,
\begin{equation}
    \begin{aligned}
        &S(\Lambda+B||\Lambda) \\&=  \left. \frac{\d}{\d R} \tr((\Lambda+B)^{R}) \right|_{R\rightarrow 1} - \left. \frac{\d}{\d R} \tr(\Lambda^{R}) \right|_{R\rightarrow 1}\\
        &= \left. \frac{\d}{\d R} {\sum_{\eta\in \mathcal{V}}  \Pr(\eta)\left[ \sum_{m \in \mathcal{M}} \sum_{l \in \mathcal{L}} \Pr(\eta+m+l)\right]^{R-1}}\right|_{R\rightarrow 1}\\& - \left.\frac{\d}{\d R} {\sum_{\eta\in \mathcal{V}}  \Pr(\eta)\left[ \sum_{m \in \mathcal{M}}  \Pr(\eta+m)\right]^{R-1}}\right|_{R\rightarrow 1}\\
        &=  \sum_{\eta\in \mathcal{V}}  \Pr(\eta) \log \frac{  \sum_{m \in \mathcal{M}} \sum_{l \in \mathcal{L}} \Pr(\eta+m+l) }{  \sum_{m \in \mathcal{M}}  \Pr(\eta+m) }.
    \end{aligned}
    \label{eq:stabilizer-RE}
\end{equation}

Now we apply the definition of Shannon conditional entropy Eq. \eqref{eq:shannon} to Eq. \eqref{eq:stabilizer-RE} and find that
\begin{equation}
\begin{aligned}
    S(\Lambda+B||\Lambda)  &=  \sum_{\eta\in \mathcal{V}}  \Pr(\eta) \log \frac{  \sum_{m \in \mathcal{M}} \sum_{l \in \mathcal{L}} \Pr(\eta+m+l) }{  \sum_{m \in \mathcal{M}}  \Pr(\eta+m) }\\
    &=  \sum_{s\in\mathcal{S}} \sum_{l'\in\mathcal{L}} \sum_{m'\in \mathcal{M}}  \Pr(s+l'+m')\\&\times \log \frac{  \sum_{m \in \mathcal{M}} \sum_{l \in \mathcal{L}} \Pr(s+l'+m'+m+l) }{  \sum_{m \in \mathcal{M}}  \Pr(s+l'+m'+m) }\\
    &=  \sum_{s\in\mathcal{S}} \sum_{l'\in\mathcal{L}} \sum_{m'\in \mathcal{M}}  \Pr(s+l'+m') \log \frac{ \Pr(s) }{ \Pr(s,l') }\\
    &=  -\sum_{s\in\mathcal{S}} \sum_{l'\in\mathcal{L}}  \Pr(s,l') \log{ \Pr(l'|s) }\\
    &=H(L|S),
\end{aligned}
\end{equation}
i.e. the AQEC relative entropy is equal to the Shannon conditional entropy of the decoding process. 

Similarly, substituting the definition of decoding SM model Eq. \eqref{eq:partition}, the AQEC relative entropy Eq. \eqref{eq:stabilizer-RE} can be expressed as
\begin{equation}
\begin{aligned}
    S(\Lambda+B||\Lambda) &= \sum_{\eta \in \mathcal{V}} \Pr(\eta) \log \frac{\sum_{l\in\mathcal{L}}Z(\eta+l)}{Z(\eta)} \\
    &= \sum_{\eta \in \mathcal{V}} \Pr(\eta) \log \left\{ \sum_{l \in \mathcal{L}} \exp\left[ -\Delta(\eta,l)\right]\right\}.
\end{aligned}
\end{equation}
This expression measures the free energy difference between the SM models with or without including the logical class d.o.f. $l$, and is a generalized version of homological difference defined in \cite{kovalevNumericalAnalyticalBounds2018}.
\end{proof}

\section{imperfect measurement prepared toric code}\label{sec:SIII}
Here we evaluate the AQEC relative entropy for the imperfect measurement prepared toric code.
The Kraus operator of single qubit Pauli $X$ error takes the form
\begin{equation}
    E_{c^*} = \sqrt{\Pr(c^*)}X_{c^*}= \sqrt{p^{|c^*|}(1-p)^{n-|c^*|}} X_{c^*},
\end{equation}
where we view Pauli $X$ error strings as cochain on the lattice \cite{Bombin-arxiv}, and $X_{c^*} = \prod_{e\in c^*} X_e$. 
In order to compute the AQEC relative entropy, we need to know about the quantity
\begin{equation}
    \bra{q_1}_L X_{c^*} \ket{q_2}_L, \quad q_i = \{++\},\{+-\},\{-+\}, \text{ or }\{--\}.
\end{equation}
Here we show the derivation. When $X_{c^*}$ is open-ended $\partial^*c^*\neq 0$, the action of $X_{c^*}$ on the post-measurement state is:
\begin{widetext}
\begin{equation}
\begin{aligned}
    X_{c^*} \ket{++}_L &= \frac{X_{c^*} M_{+} \bigotimes_e \ket{+}_e}{ \sqrt{\bra{+}^{\otimes n} M_{+}^\dagger M_{+} \ket{+}^{\otimes n}}}\\
    &= \frac{ \left(\frac{1}{\sqrt{2\cosh{\beta}}}\right)^{n/2}}{\sqrt{\bra{+}^{\otimes n} M_{+}^\dagger M_{+} \ket{+}^{\otimes n}}} \exp\left[\frac{1}{2}\beta \left(\sum_{p \notin \partial^*c^*} B_p - \sum_{p \in \partial^*c^*} B_p\right)\right] X_{c^*} \bigotimes_e \ket{+}_e\\
    &= \frac{ \left(\frac{1}{\sqrt{2\cosh{\beta}}}\right)^{n/2}}{\sqrt{\bra{+}^{\otimes n} M_{+}^\dagger M_{+} \ket{+}^{\otimes n}}} \exp\left[\frac{1}{2}\beta \left(\sum_{p \notin \partial^*c^*} B_p - \sum_{p \in \partial^*c^*} B_p\right)\right] \bigotimes_e \ket{+}_e\\
    &= \frac{1}{ \sqrt{\bra{+}^{\otimes n} M_{+}^\dagger M_{+} \ket{+}^{\otimes n}}} \exp\left[- \beta\sum_{p \in \partial^*c^*} B_p\right]M_{+} \bigotimes_e \ket{+}_e =\\& \exp\left[- \beta\sum_{p \in \partial^*c^*} B_p\right] \ket{++}_L.
\end{aligned}
\end{equation}
It changes the sign of $B_p$ operators that belong to the endpoints of $c^*$, $\partial^*c^*$, and we rewrite its action as an operator $\exp\left[- \beta\sum_{p \in \partial^*c^*} B_p\right]$ considering the fact that $B_p$ commute with each other. Notice that when $\partial^*c^*=0$, $X_{c^*}$ factories into $A_v$ stabilizers and $X$ logical operators and trivially $X_{c^*} \ket{++}_L = \ket{++}_L$.

We can expand the post-measurement state under computational (Pauli $Z$) basis,
\begin{equation}
    \exp\left[\frac{1}{2}\beta B_p\right] \ket{+}^{\otimes n} = \frac{1}{2^{n/2}}\sum_{\{\sigma_e=\pm\}} \exp\left[\frac{1}{2}\beta U_p\right] \ket{\{\sigma_e\}},
\end{equation}
where $\sigma_e$ is the eigenvalue of $Z_e$, $U_p = \prod_{e \in \partial p} \sigma_e$ is the eigenvalue of $B_p$ and we denote $Z_+ = \sum_{\{\sigma_e\}} \exp\left[\beta U_p\right]$ as the partition function of $\mathbb{Z}_2$ gauge theory.
Assume $\partial^*c^* \neq 0$, the expectation value of $X_{c^*}$ can be computed through:
\begin{equation}
\begin{aligned}
    \bra{++}_L X_{c^*} \ket{++}_L &= \bra{++}_L \exp\left[- \beta\sum_{p \in \partial^*c^*} B_p\right] \ket{++}_L\\
    &= \frac{1}{Z_+} \sum_{\{\sigma_e=\pm \}} \exp\left[\beta \sum_{p \notin \partial^*c^*} U_p\right]= \frac{2^{n/2+1}}{Z_+}  \sum_{\{U_p = \pm\}}\frac{1+\prod_p U_p}{2}   \exp\left[\beta \sum_{p \notin \partial^*c^*} U_p\right]\\
    &=\frac{2^{n/2}}{Z_+} \left(\prod_{p\notin \partial^*c^*}  \sum_{U_p}\exp\left[\beta U_p\right] \prod_{p\in \partial^*c^*} \sum_{U_p} 1 \right)+\frac{2^{n/2}}{Z_+}\left(\prod_{p\notin \partial^*c^*}  \sum_{U_p}U_p\exp\left[\beta U_p\right] \prod_{p\in \partial^*c^*} \sum_{U_p} U_p \right)\\
    &=\frac{2^{n/2}}{Z_+} (2^{n/2} (\cosh\beta)^{{n/2}-|\partial^*c^*|})\\
    &=\frac{1}{(\cosh\beta)^{|\partial^*c^*|}} \frac{1}{1+(\tanh\beta)^{n/2}},
\end{aligned}
\label{eq:pp}
\end{equation}
Here in the second line, we replaced the summation of $\sigma$ spins by the summation of $U_p = \pm$ variables. To do so, we notice that the $U_p$ variables satisfy a global constrain $\prod_{p} U_p= 1$, so we include a $\frac{1+\prod_p U_p}{2}$ factor in the summation. Counting the number of d.o.f., there are $n$ $\sigma$ spins, but only $n/2-1$ free $U_p$ variables, that is where the $2^{n/2+1}$ factor comes from.  
As for other matrix elements,
\begin{equation}
    \bra{q_1}_L X_{c^*} \ket{q_2}_L = \bra{++}_L Z_{q_1} X_{c^*} Z_{q_2} \ket{++}_L = \chi(X_{c^*}, Z_{q_2}) \bra{++}_L Z_{q_1+{q_2}} \exp\left[- \beta\sum_{p \in \partial^*c^*} B_p\right]  \ket{++}_L,
\end{equation}
where $Z_{q}$ denotes the logical $Z$ operator that send $\ket{++}_L$ to $\ket{q}_L$, for example $Z_{--} = Z_{l_1}Z_{l_2}$, $\chi(X_{c^*}, Z_{q_2})=\pm 1$ is the constant commutation factor between $X_{c^*}$ and $Z_{q_2}$.
In the above expression, the off-diagonal terms $q_1 \neq q_2$ vanishes since then $Z_{q_1+q_2}$ changes since under $1$-form symmetry operation $X_{l^*_1}$ or $X_{l^*_2}$. The diagonal value follows directly from Eq. \eqref{eq:pp} with an additional factor of the commutator,
\begin{equation}
    \bra{q_1}_L X_{c^*} \ket{q_2}_L =  \frac{1}{(\cosh\beta)^{|\partial^*c^*|}} \frac{1 + (\tanh\beta)^{n/2}\delta(\partial^* c^*=0)}{1+(\tanh\beta)^{n/2}}\chi(X_{c^*}, Z_{q_2})\delta_{q_1,q_2}.
\end{equation}

Therefore, we obtain the matrix element of $\Lambda + B$,
\begin{equation}
    \begin{aligned}
        (\Lambda + B)_{c^*_1c^*_2,q_1q_2} &= \sqrt{\Pr(c^*_1)\Pr(c^*_2)} \bra{q_1}_L X_{c^*_1}X_{c^*_2} \ket{q_2}_L \\
        &= \frac{1}{4} \sqrt{\Pr(c^*_1)\Pr(c^*_2)}  \frac{1 + (\tanh\beta)^{n/2}\delta(\partial^* c^*_1+\partial^* c^*_2=0)}{(\cosh\beta)^{|\partial^*c^*_1 + \partial^*c^*_2|}[1+(\tanh\beta)^{n/2}]} \chi(X_{c^*_1 + c^*_2}, Z_{q_2})\delta_{q_1,q_2}
    \end{aligned}
    \label{eq:matrix-element-weak1}
\end{equation}
We assign a logical $X$ operation $l^*(c^*)$ to each $c^*$, such that $[X_{l^*(c^*)},Z_q] = [X_{c^*}, Z_q]$,  $\forall q$. $l^*(c^*)$ represents the logical class of $c^*$. The Eq. \eqref{eq:matrix-element-weak1} is equivalent to 
\begin{equation}
    \begin{aligned}(\Lambda + B)_{c^*_1c^*_2,q_1q_2}= \frac{1}{4} \sqrt{\Pr(c^*_1)\Pr(c^*_2)}  \frac{1 + (\tanh\beta)^{n/2}\delta(\partial^* c^*_1+\partial^* c^*_2=0)}{(\cosh\beta)^{|\partial^*c^*_1 + \partial^*c^*_2|}[1+(\tanh\beta)^{n/2}]}\bra{q_1}X_{l^*(c^*_1)}X_{l^*(c^*_2)} \ket{q_2} \delta_{q_1,q_2}.
    \end{aligned}
    \label{eq:matrix-element-weak2}
\end{equation}
The matrix elements of $\Lambda$ does not depend on the code word $q$, thus
\begin{equation}
    \begin{aligned}\Lambda_{c^*_1c^*_2,q_1q_2}= \frac{1}{4} \sqrt{\Pr(c^*_1)\Pr(c^*_2)}  \frac{1 + (\tanh\beta)^{n/2}\delta(\partial^* c^*_1+\partial^* c^*_2=0)}{(\cosh\beta)^{|\partial^*c^*_1 + \partial^*c^*_2|}[1+(\tanh\beta)^{n/2}]} \delta_{l^*(c^*_1),l^*(c^*_2)} \delta_{q_1,q_2}.
    \end{aligned}
\end{equation}

We compute AQEC relative entropy through the replica method. To do so we need to introduce a replica index $\alpha = 1,\cdots ,R \in \mathbb{Z}_R$ to the error chain $c^{*(\alpha)}$. We alternatively represent the error configuration $c^{*(\alpha)}$ with $\mathbb{Z}_2$ spin variables $\{\eta_e^{(\alpha)} = \pm 1\}$ assigned with each edge $e$ and replica copy $\alpha$. $\eta_e^{(\alpha)} = -1$ suggests $e \in c^{*(\alpha)}$ and $\eta_e^{(\alpha)} = 1$ otherwise. The coboundary of $c^{*(\alpha)}$, $\partial^*  c^{*(\alpha)}$,is marked by $U_p^{(\alpha)} = \prod_{e \in \partial p} \eta_e^{(\alpha)}$, while $U_p^{(\alpha)} = -1$ means $p \in \partial^*  c^{*(\alpha)}$. Written in the spin variables, we have
\begin{equation}
    \begin{aligned}
        &\Pr(c^{*(\alpha)}) = \frac{\exp\left(h\sum_{e}\eta_e^{(\alpha)} \right)}{(2\cosh h)^n},\quad h = \frac{1}{2} \log \frac{1-p}{p},\\
        &|\partial^*c^{*(\alpha)} + \partial^*c^{*(\alpha+1)}| = \frac{n/2- \sum_p U_p^{(\alpha)} U_p^{(\alpha+1)}}{2},\\
        &\delta(\partial^* c^{*(\alpha)}+\partial^* c^{*(\alpha+1)}=0) = \delta_{U_p^{(\alpha)}, U_p^{(\alpha+1)}}.
    \end{aligned}
\end{equation}
Using the above substitutions, we have
\begin{equation}
    \begin{aligned}
         \tr \left( (\Lambda+B)^R\right)& = \frac{1}{4^{R-1}} \left(\prod_{\alpha=1}^R \sum_{c^{*(\alpha)}} \right) \left(\prod_{\alpha=1}^R \Pr( c^{*(\alpha)}) \right) \prod_{\alpha=1}^R \frac{1 + (\tanh\beta)^{n/2}\delta(\partial^* c^{*(\alpha)}+\partial^* c^{*(\alpha+1)}=0)}{(\cosh\beta)^{|\partial^*c^{*(\alpha)} + \partial^*c^{*(\alpha+1)}|}[1+(\tanh\beta)^{n/2}]} \\
         &= \frac{1}{4^{R-1}(2\cosh h)^{nR} (\cosh\beta)^{nR/4}(1+(\tanh \beta)^{n/2})^R} \\
         &\times \sum_{\{\eta^{(\alpha)}_e\}} \exp\left\{  \sum_{\alpha=1}^R \left[ h  \sum_{e}\eta^{(\alpha)}_e +  \frac{1}{2} \log \cosh \beta  \sum_{p}U^{(\alpha)}_p U^{(\alpha+1)}_p + \log \left(1 + (\tanh \beta)^{n/2} \prod_{p} \delta_{U^{(\alpha)}_p, U^{(\alpha+1)}_p}\right) \right]\right\}.
    \end{aligned}
\end{equation}
This is a classical partition function $Z(\beta,h) = \sum_{\{\eta^{(\alpha)}_e\}} \exp[- H(\{\eta^{(\alpha)}_e\})]$ for the Hamiltonian 
\begin{equation}
     H(\{\eta^{(\alpha)}_e\}) = - \sum_{\alpha=1}^R \left[ h  \sum_{e}\eta^{(\alpha)}_e +  \frac{1}{2} (\log \cosh \beta)  \sum_{p}U^{(\alpha)}_p U^{(\alpha+1)}_p + \log \left(1 + (\tanh \beta)^{n/2} \prod_{p} \delta_{U^{(\alpha)}_p, U^{(\alpha+1)}_p}\right) \right].
     \label{eq:hamiltonian-weak}
\end{equation}
\end{widetext}
In computing $\tr(\Lambda^R)$, we need to insert $\prod_{\alpha} \delta_{l^*(c^{*(\alpha)}),l^*(c^{*(\alpha+1)})}$ in the partition function. It forces the error configuration of each replica copy to be in the same logical class, in other words they have the same commutation relation with the $Z$ logical operators. We notice that
\begin{equation}
    \prod_{\alpha=1}^R \delta_{l^*(c^{*(\alpha)}),l^*(c^{*(\alpha+1)})} = \prod_{\alpha=1}^R \prod_{l \in \{ l_1,l_2\} } \frac{1+\prod_{e\in l} \eta^{(\alpha)}_e\eta^{(\alpha+1)}_e}{2},
\end{equation}
where $l=l_1, l_2$ labels the set of data qubits in $Z$ logical operator $Z_{l_1}, Z_{l_2}$.
The R\'enyi version of AQEC relative entropy can be represented by the classical expectation value of the above expression,
\begin{equation}
\begin{aligned}
        S^{(R)}(\Lambda+B||\Lambda) &= \frac{1}{R-1} \log \frac{\tr((\Lambda+B)^R)}{\tr(\Lambda^{R})}\\& = \frac{1}{1-R} \log \Braket{\prod_{\alpha,l} \frac{1+\prod_{e\in l} \eta^{(\alpha)}_e\eta^{(\alpha+1)}_e}{2}},
\end{aligned}
\label{eq:entropy-weak}
\end{equation}
where we have used $\tr_{\mathcal{C}} (B) = 0$.
The SM model Eq. \eqref{eq:hamiltonian-weak} captures the fluctuation of error configurations $\{\eta^{(\alpha)}_e\}$. The decodable order corresponds to that all spins are aligned up, $eta^{(\alpha)}_e=+$. 
The first term tends to point the spins up, while the second and third terms tend to pin the endpoints of the error chains on different replica copies together at the same plaquette. 
Large temperatures or small Interaction parameters $\beta$, $h$ introduce disorder to the SM system. 

We now discuss the extremal cases of Eq. \eqref{eq:entropy-weak}.
We first analyze the low-temperature limit.
Taking the $\beta \rightarrow +\infty$ limit and keeps $h$ finite, the partition function becomes
\begin{equation}
    Z(\beta,h) \propto \sum_{\{\eta^{(\alpha)}_e\}} \left(\prod_{\alpha, p} \delta_{U^{(\alpha)}_p, U^{(\alpha+1)}_p } \right) \exp \left[ h\sum_{\alpha, e} \eta^{(\alpha)}_e \right],
\end{equation}
with the constraint that all endpoints of the replica copies are exactly at the same plaquettes. Following the same spirit in deriving Eq. \eqref{eq:trLB}, $\partial^* c^{*(\alpha)}=\partial^* c^{*(\alpha+1)}$ for all $\alpha$ suggests that we can rewrite $c^{*(\alpha)} =  c^{*(R)} + b^{*(\alpha)}+l^{*(\alpha)}$ where $b^{*(\alpha)}$ is a coboundary (contractible loops) and $l^{*(\alpha)}$ is a cohomological class (non-contractible loops or logical class). We can further rewrite the summation of coboundaries as summations of spin variables on vertices $\sigma_{v}^{(\alpha)}$ by viewing $b^{*(\alpha)}$ as domain walls. Sign difference of $\sigma$ spins at two ends of a edges $\prod_{v\in \partial e}\sigma_{v}^{(\alpha)}=-1$ corresponds to $\{\eta^{(\alpha)}_e\}=-1$. We also denote $\eta^{l^{*(\alpha)}}_e=-1$ when $e\in l^{*(\alpha)}$ and $\eta^{l^{*(\alpha)}}_e=+1$ otherwise. Thus
\begin{equation}
\begin{aligned}
    Z(\beta,h) &\propto \sum_{\{\eta^{(R)}_e\}}\sum_{\{\sigma^{(\alpha)}_v\}} \sum_{\{l^{*(\alpha)}\}} \exp \left[ h\sum_{ e} \eta^{(R)}_e  \right.\\ &\left.+ h\sum_{\alpha=1}^{R-1} \sum_{ e} \eta^{(R)}_e \eta^{l^{*(\alpha)}}_e \prod_{v\in \partial e}\sigma_{v}^{(\alpha)} \right],
\end{aligned}    
    \label{eq:RBIM-L}
\end{equation}
and self-consistently arrive at the replica partition function of the random bond Ising model (RBIM) \cite{nishimoriStatisticalPhysicsSpin2001,ref:dennis} with additional fluctuation of non-contractible domain walls. 
When computing $\tr\left(\Lambda^R\right)$, the non-contractible fluctuations are forbidden and we obtain 
\begin{equation}
\begin{aligned}
    &\tr\left(\Lambda^R\right) \\&\propto \sum_{\{\eta^{(R)}_e\}}\sum_{\{\sigma^{(\alpha)}_v\}}\exp \left[ h\sum_{ e} \eta^{(R)}_e + h\sum_{\alpha=1}^{R-1} \sum_{ e} \eta^{(R)}_e \prod_{v\in \partial e}\sigma_{v}^{(\alpha)} \right] \\&\propto \sum_{\{\eta^{(R)}_e\}} \Pr(\{\eta^{(R)}_e\}) Z_{RBIM}(\{\eta^{(R)}_e\})^{R-1},
\end{aligned}
    \label{eq:RBIM}
\end{equation}
It means that at the axis $\mathcal{T} = 1/\beta = 0$, the intrinsic threshold is located at $p_{th} \simeq 0.11$ \cite{ref:dennis}. Note that the phase transition point might be different for different $R$ \cite{ashvin} and the correct value of threshold is obtained by taking the $R\rightarrow 1$ limit. 
When $h\rightarrow +\infty$ and fixing $\beta$, all $\eta$ spins are forced to point up $\{\eta^{(\alpha)}_e\}=+1$, and the AQEC relative entropy is trivially $0$.

Then we analyze the high-temperature limit. Take $\beta \rightarrow 0$, the SM model becomes a trivial paramagnetic model, 
\begin{equation}
    Z(\beta,h) = \sum_{\{\eta^{(\alpha)}_e\}}  \exp \left[ h\sum_{\alpha, e} \eta^{(\alpha)}_e \right] = 2^n \cosh^n h,
\end{equation}
Assume the lattice is square with linear size $L = \sqrt{n/2}$, which is also the weight of logical operators. Through a straightforward calculation of the expectation value in Eq. \eqref{eq:entropy-weak} with the above paramagnetic model, we get
\begin{equation}
    S^{(R)}(\Lambda+B||\Lambda) = \left[\left(\frac{1+\tanh^L h}{2}\right)^R+\left(\frac{1-\tanh^L h}{2}\right)^R\right]^2.
\end{equation}
Take the thermodynamic limit $n\rightarrow \infty$ and the replica limit $R\rightarrow 1$, we have
\begin{equation}
    \lim_{R\rightarrow 1} \lim_{n \rightarrow \infty} S^{(R)}(\Lambda+B||\Lambda) = 2\log 2.
\end{equation}
The reason why it does not saturate the upper bound $4\log 2$ is that we only considered Pauli $X$ noises and ignored Pauli $Z$ noises. In the $h \rightarrow 0$ limit with finite $\beta$, the Hamiltonian has the form 
\begin{equation}
\begin{aligned}
    &H(\{\eta^{(\alpha)}_e\}) = - \sum_{\alpha=1}^R \left[  \frac{1}{2} (\log \cosh \beta)  \sum_{p}U^{(\alpha)}_p U^{(\alpha+1)}_p \right.\\&\left. + \log \left(1 + (\tanh \beta)^{n/2} \prod_{p} \delta_{U^{(\alpha)}_p, U^{(\alpha+1)}_p}\right) \right].
\end{aligned}
\end{equation}
Ignoring the last term, the model is basically a union of one-dimensional Ising models of the $U_{p}^{(\alpha)} = \pm 1$ variables along the replica direction. Since the $1$-D Ising models are disordered at finite temperatures, we anticipate that they lead to undecodable phase. 

We then consider the monotonicity of $S^{(R)}(\Lambda+B||\Lambda)$ varying interaction parameters $h$, $\beta$.
For a finite $\beta \geq 0$, the third term of Eq. \eqref{eq:hamiltonian-weak} is exponentially suppressed and negligible in the thermodynamic limit $n\rightarrow \infty$, and we approximate the Hamiltonian by
\begin{equation}
\begin{aligned}
     &H(\{\eta^{(\alpha)}_e\}) = - \sum_{\alpha=1}^R \left[ h  \sum_{e}\eta^{(\alpha)}_e +  J  \sum_{p}U^{(\alpha)}_p U^{(\alpha+1)}_p \right], \\& J=\frac{1}{2} \log \cosh \beta
\end{aligned}
     \label{eq:hamiltonian-weak2}
\end{equation}
For a ferromagnetic spin model $J \geq 0$, the Griffiths-Kelly-Sherman (GKS) inequalities holds \cite{Griffiths,Kelly,jiangDualityFreeEnergy2019}, 
\begin{equation}
    \braket{\Gamma_\mathcal{A}} \geq 0, \quad \braket{\Gamma_\mathcal{A} \Gamma_\mathcal{B}} \geq \braket{\Gamma_\mathcal{A}}\braket{\Gamma_\mathcal{B}},
\end{equation}
where $\mathcal{A}$, $\mathcal{B}$ denotes sets containing pairs of edge and replica copy, $(e,\alpha)$, and $\Gamma_\mathcal{A}$, $\Gamma_\mathcal{B}$ is the product of corresponding spin variables, 
\begin{equation}
    \Gamma_X = \prod_{(e,\alpha) \in \mathcal{X}} \eta_e^{(\alpha)}, \quad X = \mathcal{A},\mathcal{B}.
\end{equation}
In particular, we have
\begin{equation}
\begin{aligned}
    &    \frac{\d}{\d J} \braket{\Gamma_\mathcal{A}} = \sum_p \left\{\Braket{\Gamma_\mathcal{A}U^{(\alpha)}_p U^{(\alpha+1)}_p} - \Braket{\Gamma_\mathcal{A}} \Braket{U^{(\alpha)}_p U^{(\alpha+1)}_p} \right\} \geq 0,\\
    &     \frac{\d}{\d h} \braket{\Gamma_\mathcal{A}} = \sum_e \left\{\Braket{\Gamma_\mathcal{A}\eta^{(\alpha)}_e} - \Braket{\Gamma_\mathcal{A}} \Braket{\eta^{(\alpha)}_e} \right\} \geq 0,
\end{aligned}
\end{equation}
Notice that the expectation value in Eq. \eqref{eq:entropy-weak} can be expanded into summation of terms like $\Gamma_{\mathcal{A}}$ for different sets of spin variables $\mathcal{A}$, so we conclude that 
\begin{equation}
\begin{aligned}
    &\frac{\d}{\d J}  \Braket{\prod_{\alpha,l} \frac{1+\prod_{e\in l} \eta^{(\alpha)}_e\eta^{(\alpha+1)}_e}{2}} \geq 0, \\& \frac{\d}{\d h}  \Braket{\prod_{\alpha,l} \frac{1+\prod_{e\in l} \eta^{(\alpha)}_e\eta^{(\alpha+1)}_e}{2}} \geq 0.
\end{aligned}
\end{equation}
In the thermodynamic limit, the AQEC relative entropy should be monotonically increasing with $\beta$ and $h$ following the above discussions, 
\begin{equation}
\begin{aligned}
   &\left. \frac{\d}{\d \beta}  S^{(R)}(\Lambda+B||\Lambda) \right|_{n\rightarrow \infty}\geq 0, \\& \left. \frac{\d}{\d h}  S^{(R)}(\Lambda+B||\Lambda) \right|_{n\rightarrow \infty} \geq 0.
\end{aligned}
   \label{eq:mono}
\end{equation}
Note that $S^{(R)}(\Lambda+B||\Lambda)$ might be discontinuous at the phase transition point in the $n\rightarrow \infty$ limit, and we keep in mind that the above inequality includes the case that the derivative behaves like a delta function.

Now in order to study the properties of AQEC relative entropy at low but finite temperatures, we assume $e^{-\beta} \sim e^{-h} \ll 1$ and expand Eq. \eqref{eq:entropy-weak} to the first non-vanishing order. Recall that the quantity $\prod_{\alpha,l} \frac{1+\prod_{e\in l} \eta^{(\alpha)}_e\eta^{(\alpha+1)}_e}{2}$ measures the difference of logical classes of the replica copies. It is $1$ when all copies have the same logical class and $0$ when there is a difference. In low temperature, the ground state is that all spins stay in $+1$, and the lowest order excitation is a single spin flip $\eta^{(\alpha)}_e=-1$ which costs energy $2h + 8J +2 \log (1+ \tanh^{n/2} \beta)$.
Here the $2h$ term comes from the magnetic field term $-h \sum_{p,\alpha} \eta^{(\alpha)}_e$. The $8J$ term comes from the interaction $-J \sum_{p,\alpha}U^{(\alpha)}_p U^{(\alpha+1)}_p$. Note that in the interaction term, each $\eta^{(\alpha)}_e$ spin is involved in two replica copies, and in each copy it is involved in $2$ plaquettes, leading to the energy increment $4\times 2J$. The $2 \log (1+ \tanh^{n/2} \beta)$ term comes from the last term of the Hamiltonian Eq. \eqref{eq:hamiltonian-weak}. There are $nR$ possible configurations of such excitation, so the partition function reads
\begin{equation}
\begin{aligned}
    Z(\beta,h) &\simeq 1+ nR \exp [-2h - 8J -2 \log (1+ \tanh^{n/2} \beta)] \\&\simeq 1 + 4nR e^{-2h - 4\beta},
\end{aligned}
\end{equation}
normalized by the ground state Boltzmann weight.
Notice that We have substituted $J= (1/2)\log \cosh \beta$ and expanded up to the order $e^{-4\beta}$. In this case both the $8J$ term and $2 \log (1+ \tanh^{n/2} \beta)$ term contribute to the factor $4$.
Now we insert $\prod_{\alpha,l} \frac{1+\prod_{e\in l} \eta^{(\alpha)}_e\eta^{(\alpha+1)}_e}{2}$ in the summation, and it yields $0$ When the flipped spin $\eta^{(\alpha)}_e=-1$ is located on $Z$ logical operators $l_1$, $l_2$. There are $2L R$ such configurations, thus 
\begin{equation}
\begin{aligned}
  &\sum_{\{\eta^{(\alpha)}_e\}} \prod_{\alpha,l} \frac{1+\prod_{e\in l} \eta^{(\alpha)}_e\eta^{(\alpha+1)}_e}{2} \exp[-H] \\&\simeq 1+ (nR-2LR) \exp [-2h - 8J -2 \log (1+ \tanh^{n/2} \beta)] \\&\simeq 1 + (4n-8L)R e^{-2h - 4\beta},    
\end{aligned}
\end{equation}
So the expectation value is
\begin{equation}
\begin{aligned}
    &\Braket{\prod_{\alpha,l} \frac{1+\prod_{e\in l} \eta^{(\alpha)}_e\eta^{(\alpha+1)}_e}{2}} \\&\simeq \frac{1 + (4n-8L)R e^{-2h - 4\beta}}{1 + 4nR e^{-2h - 4\beta}} \\&\simeq 1 - 8LR e^{-2h - 4\beta}.
\end{aligned}
\end{equation}
Substitute in Eq. \eqref{eq:entropy-weak}, we can approximate the R\'enyi AQEC relative entropy by
\begin{equation}
    S^{(R)}(\Lambda+B||\Lambda) \simeq \frac{8LR}{R-1} e^{-2h-4\beta}.
    \label{eq:expansion-app}
\end{equation}

Although there is a subtlety in this expression that the $R\rightarrow 1$ and $n\rightarrow \infty$ limits are not compatible with the perturbative expansion, it nonetheless suggests that the imperfect code cannot suppress $S^{(R)}(\Lambda+B||\Lambda)$ to $0$ and lead to a phase transition at $1/\beta=0$ for any $R>1$.
In the $n\rightarrow\infty$ limit where the third term of the Hamiltonian is negligible, $S^{(R)}(\Lambda+B||\Lambda)$ is monotonically increasing with $1/\beta$ as in Eq. \eqref{eq:mono}. The divergent coefficient $L$ in Eq. \eqref{eq:expansion-app} indicates a sudden jump at $e^{-4\beta} \rightarrow 0$. Keep $e^{-h}$ small but finite, we have 
\begin{equation}
    \left. \frac{\d}{\d e^{-4\beta}}  S^{(R)}(\Lambda+B||\Lambda) \right|_{e^{-4\beta} \rightarrow 0+} = \frac{8LR}{R-1} e^{-2h}.
\end{equation}
It diverges in the thermodynamic limit,
\begin{equation}
    \lim_{n\rightarrow \infty}\left. \frac{\d}{\d e^{-4\beta}}  S^{(R)}(\Lambda+B||\Lambda) \right|_{e^{-4\beta} \rightarrow 0+} \rightarrow +\infty.
\end{equation}
Similarly keep $e^{-\beta}$ small but finite, we have
\begin{equation}
\begin{aligned}
    &\left. \frac{\d}{\d e^{-2h}}  S^{(R)}(\Lambda+B||\Lambda) \right|_{e^{-2h} \rightarrow 0+} = \frac{8LR}{R-1} e^{-4\beta},\\& \lim_{n\rightarrow \infty}\left. \frac{\d}{\d e^{-2h}}  S^{(R)}(\Lambda+B||\Lambda)  \right|_{e^{-2h} \rightarrow 0+} \rightarrow +\infty.
\end{aligned}
\end{equation}
So we conclude that the phase diagram has the form in Fig. \ref{fig3} for any $R>1$, as well as $R\rightarrow 1$ after performing extrapolation.
The system stays in undecodable phase as long as the preparation and Pauli error rates are both finite.  The intuition for this phenomenon is gained from the SM model. When the preparation is perfect $\beta \rightarrow +\infty$, then endpoints of different replica copies are pinned together and it is hard to create logical difference between replica copies.

\end{appendix}

\bibliographystyle{apsrev4-1-titles} 
\bibliography{ref}

\end{document}